\documentclass[12pt,a4paper]{article}

\usepackage[margin=3cm]{geometry}

\usepackage{setspace}
\usepackage{graphicx}
\usepackage{amsmath, amssymb}
\usepackage{hyperref}
\usepackage{natbib}
\usepackage{float}

\usepackage{multicol}
\usepackage{mathtools}
\usepackage[labelfont=bf]{caption}
\usepackage{subfigure}
\begin{document}
\begin{center}
\large \textbf{Bayesian model averaging in model-based clustering and density estimation}\\[12pt]

\normalsize Niamh Russell$^{1, 2\ast}$, Thomas Brendan Murphy$^{1,3}$, Adrian E. Raftery$^4$ \\[12pt]
	$^1$ School of Mathematical Sciences, University College Dublin, Ireland\\
$^2$Complex and Adaptive Systems Laboratory, UCD, Ireland.\\ 
$^3$Insight Research Centre, UCD, Ireland.\\ 
$^4$Department of Statistics, University of Washington, Seattle, Washington, USA\\

	$^{\ast}$Email: niamh.russell.1@ucdconnect.ie\\[24pt]
\end{center}



%
%
%
%
%
{\small \textbf{Abstract:}
We propose Bayesian model averaging (BMA) as a method for postprocessing the results of model-based clustering. Given a number of competing models, appropriate model summaries are averaged, using the posterior model probabilities, instead of being taken from a single ``best'' model. We demonstrate the use of BMA in model-based clustering for a number of datasets. We show that BMA provides a useful summary of the clustering of observations while taking model uncertainty into account. Further, we show that BMA in conjunction with model-based clustering gives a competitive method for density estimation in a multivariate setting. Applying BMA in the model-based context is fast and can give enhanced modeling performance. 
}

\normalsize
\section{Introduction} 
\label{sec:Ch4intro}
Model-based clustering methods are based on the assumption that the population can be modeled using the finite mixture model \citep[e.g.][]{banfield1993, celeux1995, fraley2002}.  Within this paradigm, it is assumed that the data come from $G$ subpopulations, which correspond to the mixture components, and within each subpopulation the data are modeled using a single parametric component distribution. The most common finite mixture model that is used for model-based clustering is the finite normal mixture, but many alternatives exist \citep[e.g.][]{mclachlan2000,lee2013}.

Model selection is an intrinsic part of model-based clustering. 
In particular, the number of clusters (component densities), $G$, 
is unknown and a number of competing choices for component densities may also be under consideration. Each combination of component density and number of clusters
can be viewed as a separate model, and a model selection approach can be
used to select both at the same time \citep[e.g.][]{fraley2002}.
In most implementations of model-based clustering, the ``best" model is chosen by using some criterion and clustering is based on the ``best'' single model.  A number of methods have been proposed for selecting the ``best'' model including choosing the model with the highest Bayesian Information Criterion (BIC) \citep{schwarz1978} or the highest Integrated Complete Likelihood (ICL) \citep{biernacki2000}. 

The approach of reporting the results of model-based clustering based on a single model ignores the uncertainty that arises from the model selection. 
Consequently, the uncertainty about quantities of interest may be 
underestimated.
We propose basing the results of model-based clustering on a combination of the results from all candidate models rather than on those from a single model. 
We propose taking a weighted average of the model summaries, where the weights are approximate posterior model probabilities. 
Thus we propose using Bayesian Model Averaging (BMA) \citep{hoeting1999} within the model-based clustering paradigm. 

To obtain valid inference using BMA, the quantity of 
interest should have the same meaning in each model under consideration. 
In model-based clustering, the quantities of interest must have the same meaning for all values of $G$ and must be invariant to the labeling of the clusters in the finite mixture model. This is because finite mixture models are identifiable only up to permutations of the cluster labels.
Here we focus on inference about the clustering consensus matrix. 
This has the same meaning for all values of $G$ 
and is invariant to the cluster labeling.  

We also consider using model-based clustering as a method for 
multivariate density estimation, following \cite{fraley2002}.
In this case, the estimated density has the same meaning in all models, so we again use the posterior model probabilities as weights to offer an alternative density estimation procedure to multivariate kernel density estimation \citep[e.g.,][]{scott1992,duong2007} or to model-based clustering density estimation methods based on a single model \citep{fraley2002}. 

The paper is organized as follows.  In Section~\ref{sec:Ch4background}, 
we briefly review the model-based clustering paradigm.
In Section~\ref{sec:Ch4BMA}, we outline how the BMA approach can be used to deal with model uncertainty.   In Section~\ref{sec:Ch4method}, we describe how we can create a matrix for each model that is invariant to the number and labeling of the clusters. In Section~\ref{sec:Ch4results}, we provide an assessment of clustering performance for BMA in conjunction with model-based clustering. 
In Section~\ref{sec:Ch4density}, we describe the background for density estimation and introduce a framework for combining BMA and model-based clustering for multivariate density estimation. We illustrate this with a number of simulations. 
We conclude, in Section~\ref{sec:Ch4discuss}, with a discussion of related work and suggest future directions.

\section{Model-based clustering}
\label{sec:Ch4background}

Model-based clustering \citep{banfield1993, celeux1995,fraley2002,fraley2007}  is used for clustering data into groups, where the number of groups $G$ is typically unknown. Here we focus on model-based clustering based on the finite normal mixture model as described in \cite{fraley2002}.

We assume that there are $G$ clusters, where each cluster $g$ arises with probability $\tau_g$. Data within each cluster follow a normal distribution with cluster-specific mean $\mu_g$ and covariance $\boldsymbol{\Sigma}_g$, so that the data are characterized by a finite mixture of normal distributions. 
The density of each observation $x_i$ is
\begin{equation*}
f(x_i)=\sum^G_{g=1}\tau_g\phi(x_i|\mu_g,\boldsymbol{\Sigma}_g),
\end{equation*} 
where $\phi(\cdot|\cdot,\cdot)$ is a multivariate normal density.
In practice, the number of clusters, $G$, is usually unknown and needs to be determined as part of the model inference. 

The assumption of multivariate normal distributed clusters implies that the clusters are elliptical in shape. \cite{banfield1993} proposed that constraints be placed on the covariance matrices to gain parsimony.
These are specified using a modified eigenvalue decomposition of $\boldsymbol{\Sigma}_g$, namely
\begin{displaymath}
\boldsymbol{\Sigma}_g=\lambda_g \mathbf{D}_g \mathbf{A}_g \mathbf{D}_g^{\intercal},
\end{displaymath}
where $\lambda_g$ is a constant which controls the cluster volume,
$\mathbf{D}_g$ is an orthogonal matrix of eigenvectors which control the
orientation of the clusters, and $\mathbf{A}_g$  a diagonal matrix, 
with entries proportional to the eigenvalues, 
which control the shape of the cluster.

\begin{figure}
\begin{center}
\includegraphics[width=0.6\textwidth]{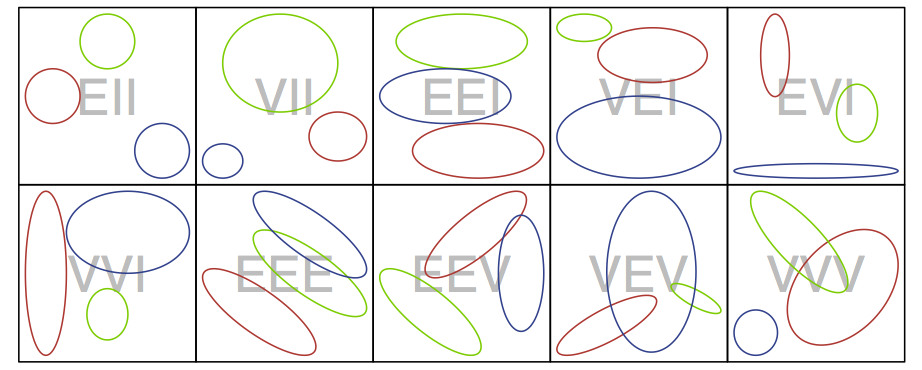}
\caption[Examples of cluster shapes under each of the first ten covariance restrictions]{Examples of cluster shapes under each covariance restriction.}
\label{fig:Ch4ellipses}
\end{center}
\end{figure}

We can restrict  each property of the covariance $\boldsymbol{\Sigma}_g$ 
(volume, shape, orientation)  in different ways, resulting in fourteen different possible models \citep{biernacki2006}. Throughout this paper, we will consider the ten covariance structures implemented in the {\sf mclust} software \citep{fraley2012}, as displayed in Figure~\ref{fig:Ch4ellipses}. Each letter in the name of a model corresponds to the constraint
placed on the volume, shape and orientation respectively. 
The constraint can be equal (E), variable (V) or identity (I). 
For example, in the EEV model, each cluster has the same volume 
and the same shape but the orientations of the clusters can differ. 
Other parametrizations of covariance matrices are useful in the context of model-based cluster analysis \citep[e.g.][]{mcnicholas2008,mcnicholas2010b,biernacki2014}, but we do not consider them further here.

Based on choosing a covariance constraint and the number of groups $G$, we can fit the finite mixture of normal distributions using the EM algorithm \citep{dempster1977} to yield the maximum likelihood estimates of the model parameters and an estimate of the posterior probability that each observation $x_i$ belongs to group $g$. The resulting $N \times G$ matrix of probabilities, denoted by $\mathbf{Z}$, has typical entry $z_{ig}$, which is the estimated conditional posterior probability that observation $i$ belongs to group $g$. This is
\begin{equation}
{\mathbb P}\{\mbox{Observation $i$ comes from group $g$}\}\approx z_{ig}=\frac{\hat\tau_g\phi(\mathbf{x}_i|\hat\mu_g,\hat{\boldsymbol\Sigma}_g)}{\sum_{g'=1}^G \hat\tau_{g'} \phi(\mathbf{x}_i|\hat\mu_{g'},\hat{\boldsymbol\Sigma}_{g'})}, \label{eq:Ch4clustpostprob}
\end{equation}
where $\{(\hat\tau_g,\hat\mu_g,\hat{\boldsymbol\Sigma}_g):g=1,2\ldots,G\}$ are the maximum likelihood estimates of the model parameters. The probability matrix, $\mathbf{Z}$, can be converted into group labels for the observations and thus the observations can be clustered.

In a clustering problem where several possible values for $G$ are considered (e.g.~$G=1,2,\ldots, 9$) and when all ten model types are also considered, we have to choose between about 83 different models. However, we would expect only a small number of these models to be strongly supported by the data. A frequently-used model selection approach is to choose the ``best" model using the Bayesian Information Criterion (BIC), \citep{schwarz1978} where the BIC for model ${\cal M}_m$ is given by
\begin{equation}
\textrm{BIC}_m=2 \log (\mathcal{L})-\kappa_m \log (N).
\label{eq-BIC}
\end{equation}
In Equation~\ref{eq-BIC}, $\mathcal{L}$ is the maximized likelihood of the data, $\kappa_m$ is the number of estimated model parameters for model ${\cal M}_m$, and $N$ is the number of observations. 
This approach was proposed for clustering by \cite{dasgupta1998},
and has been found to perform well \citep[e.g.][]{steele2010}.

Once the model has been chosen, the cluster membership matrix is based on the parameters for that model alone. The other competing models are then 
discarded and model uncertainty is ignored.

\section{Model Uncertainty and Bayesian model averaging}
\label{sec:Ch4BMA}

Basing inferences on a single ``best'' model ignores uncertainty about 
what the best model is.  This can result in underestimating uncertainty about 
quantities of interest such as cluster membership or the estimated density. 

We address this using Bayesian model averaging (BMA) 
\citep{leamer1978,madigan1994,hoeting1999}, which proceeds as follows.
If $\{{\cal M}_1,...,{\cal M}_M\}$ denotes the set of all models being considered and if $\Delta$ is the quantity of interest, then the posterior distribution of $\Delta$ given the data is
\begin{equation}
\mathbb{P}(\Delta|\text{Data})=\sum_{m=1}^{M}\mathbb{P}(\Delta|{\cal M}_m,\text{Data})\mathbb{P}({\cal M}_m|\text{Data}).
\label{eq:Ch4intro}
\end{equation}
This is an average of the posterior distributions under each model weighted by the corresponding posterior model probabilities. 

In Equation~\ref{eq:Ch4intro}, the posterior probability of model ${\cal M}_m$ is given by
\begin{equation*}
\mathbb{P}({\cal M}_m|\text{Data})=\frac{\mathbb{P}(\text{Data}|{\cal M}_m)\mathbb{P}({\cal M}_m)}{\sum_{m'=1}^M \mathbb{P}(\text{Data}|{\cal M}_{m'})\mathbb{P}({\cal M}_{m'})},
\end{equation*}
where
\begin{equation}
\mathbb{P}(\text{Data}|{\cal M}_m)=\int \mathbb{P}(\text{Data}|\theta_m,{\cal M}_m)\mathbb{P}(\theta_m|{\cal M}_m)d\theta_m
\label{eq:Ch4integral}
\end{equation}
is the marginal likelihood of model ${\cal M}_m$, $\theta_m$ is the vector of parameters of model ${\cal M}_m$, $\mathbb{P}(\text{Data}|\theta_m,{\cal M}_m)$ is the likelihood for model ${\cal M}_m$, $\mathbb{P}(\theta_k|{\cal M}_k)$ is the prior density of the parameter $\theta_m$ in model ${\cal M}_m$  and $\mathbb{P}({\cal M}_m)$ is the prior probability of model ${\cal M}_m$. All probabilities are conditional on the set of all models being considered.
\cite{madigan1994} argued that averaging over all of the models, as in Equation~\ref{eq:Ch4intro}, provides better predictive ability than using any single model. 
 
One difficulty is that the integral in Equation~\ref{eq:Ch4integral} is intractable in most cases and so an approximation to the posterior model probability is required. 
We use the Bayesian Information Criterion (BIC) to derive approximate posterior model probabilities for Bayesian model averaging for model-based clustering
\citep{dasgupta1998}.
If all the models are equally likely \emph{a priori}, so that $\mathbb{P}({\cal M}_1)=\dots =\mathbb{P}({\cal M}_M)=1/M$, this yields
\begin{equation}
\mathbb{P}({\cal M}_m|\text{Data}) \approx \frac{\exp \left(\frac{1}{2}BIC_m\right)}{\sum_{m'=1}^M\exp \left(\frac{1}{2}BIC_{m'}\right)}.
\label{eq:Ch4PMP}
\end{equation}




\section{Bayesian model averaging for clustering}
\label{sec:Ch4method}
An important point when implementing BMA is that the quantity of interest, $\Delta$, must have a consistent definition across all models. Care should be taken when implementing BMA in the clustering setting because the labeling of clusters (components) within a model is arbitrary, so the labeling of clusters under one model may not correspond to the labeling under another model. Also, the number of clusters can vary from model to model.

\subsection{Similarity matrix}
\citet{strehl2003} introduced the concept of a binary \emph{similarity matrix} in the context of clustering. This matrix is $N\times N$ and 
for any pair of observations $(i,j)$, 
the $(i,j)$th entry in the matrix is 1 if the $i$th and $j$th observations are in the same cluster in a given model, and 0 if they are not. 
They used this structure to form what they call \emph{pairwise cluster ensembles}; \cite{monti2003} and \cite{kuncheva2004} also exploited this idea in a clustering context. 
In these articles, weights were given to various clustering methods and the similarity matrices were combined using these weights. The resulting matrix was described by \citet{kuncheva2004} as a \emph{consensus matrix}. 

\citet{fern2003} extended the binary similarity matrix to soft-clustering techniques, which return a probability vector $\mathbb{P}(g|i,{\cal M}_m), g=1,\dots,G$ for an observation $i$, representing the probability that $i$ belongs to each cluster under model ${\cal M}_m$. The values ${\mathbb P}(g|i,{{\cal M}_m})$ are analogous to the $z_{ig}$ values, as defined in Equation~\ref{eq:Ch4clustpostprob}. \cite{fern2003} used model-based clustering to cluster high-dimensional data by randomly projecting the data into a low-dimensional space and clustering the data in the low-dimensional space. The data are projected multiple times and the consensus matrices are averaged across all projections.

\cite{fern2003} defined $\mathbf{S}^{{m}}_{ij}$ as the probability that observations $i$ and $j$ belong to the same cluster under model ${{\cal M}_m}$.
This can be calculated as 
\begin{equation*}
\mathbf{S}^{{m}}_{ij}=\sum_{g=1}^G \mathbb{P}(g|i,{{\cal M}_m})\times \mathbb{P}(g|j,{{\cal M}_m})\approx\sum_{g=1}^{G}z_{ig}^m z_{jg}^m ,
\end{equation*}
because $z_{ig}^m$ is an estimate of the probability that observation $i$ belongs to cluster $g$ in model ${\cal M}_m$. 

We can construct the similarity matrix $\mathbf{S}^m$ for each model ${\cal M}_m$ as follows:
\begin{equation}
\mathbf{S}_{ij}^{m} = 
\left\{
\begin{array}{ll}
(\mathbf{Z}^m(\mathbf{Z}^{m})^{\intercal})_{ij} & , \mbox{ when } i \ne j\\
1 & , \mbox{ when }  i=j . 
\end{array}
\right.
\label{eq:Ch4zzt}
\end{equation}
We propose using the matrix $\mathbf{S}^{m}$ of probabilities  that any pair of observations belong to the same cluster, when implementing BMA for model-based clustering. This ensures that we are averaging a quantity that has the same meaning across differing number of clusters and is invariant to cluster labeling.

\subsection{Properties of the similarity matrix}
\label{sec:toy}

The matrix $\mathbf{S}^m$ for any model ${\cal M}_m$ will be $N \times N$ where $N$ is the number of observations. 
It is invariant to label switching, and its dimension is invariant to the number of clusters in the model. 
It can therefore be used to combine models with different numbers of clusters. 
It can be viewed as a similarity matrix between the data points $(x_1,\dots ,x_N)$. 

The element $s_{ij}^k $ of the matrix $\mathbf{S}^m$ represents the probability that $i$ and $j$ belong to the same cluster. 
We can also define a quantity $d_{ij}=1-s_{ij}$ as the probability that they are not in the same cluster. So, $\mathbf{S} = 1-\mathbf{D}$, where $\mathbf{D}$ 
can be thought of as a dissimilarity matrix.  
Therefore, $\mathbf{D}$ can be used with any clustering algorithm 
which operates directly on a dissimilarity matrix. 

We define $s_{\mathcal{A}}$ as the minimum probability that two observations $i$ and $j$ in  a set $\mathcal{A}$ belong together, which we can think of as the minimum probability that two elements in $\mathcal{A}$ belong to the same cluster. If we define $d_{\mathcal{A}}$ as $1-s_ {\mathcal{A}}$, we obtain the following result:
\begin{eqnarray}
d_{\mathcal{A}}&=& 1-s_{\cal A}\nonumber\\
& = & 1 -  \min_{i,j \in \mathcal{A}} \mathbb{P}\{i, j \text{ belong together}\}\nonumber\\
&=& \max_{i,j \in \mathcal{A}} [1-\mathbb{P}\{i, j \text{ belong together}\}]\nonumber\\ 
&=& \max_{i,j\in\mathcal{A}}(d_{ij})\nonumber.
\end{eqnarray}

Thus the maximum probability that two elements of ${\cal A}$ do not appear in the same cluster is $d_{\cal A}$. It follows that ${\mathbf D}$ can be used in hierarchical clustering with complete linkage \citep{sokal1963} and the results will have an intuitive interpretation. In complete linkage, the dissimilarity between two groups $G$ and $H$ is the largest dissimilarity between opposite groups,
that is, the maximum distance between an element of $G$ and an element of $H$,
namely
$$
d_{\text{complete}}(G,H)= \max_{i\in G,j\in H} d_{ij}.
$$
Hierarchical complete-linkage clustering will then merge the groups with the smallest $d_{\text{complete}}$. This continues until all the groups are merged.

The results can be shown on a dendrogram which can be cut at any level. 
There is an intuitive interpretation of the level of cut. If we cut the dendrogram at a particular probability level, any observations clustered together at that level have a probability of at least that value of all being in the same cluster.

\begin{figure}
\centering
\includegraphics[width=0.8\textwidth]{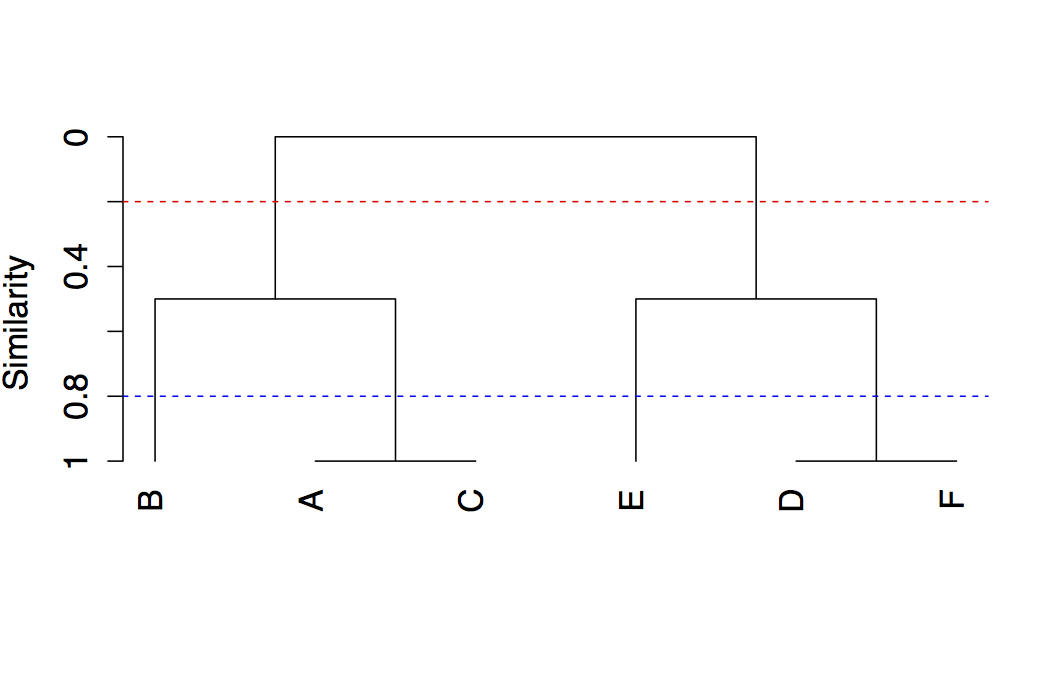}
\caption[Toy example of a dendrogram which denotes the probability of certain observations being in the same group.]{Toy data: Dendrogram denotes the probability of certain observations being in the same group. For example if we cut the dendrogram at the blue line, any observations clustered together have a probability of at least 0.8 of all being in the same group. If we cut at the red line, any observations clustered together at that level have a probability of at least 0.2 of all being in the same group.}  
\label{fig:Ch4toyHeat}
\end{figure}

We illustrate this with a toy example. Suppose we have six data points $A,\dots,F$ and two equally likely clustering results.
The first clustering attempt puts $A,B,C$ in one cluster and $D,E,F$ in the other, while the second attempt puts $A,C,E$ in one cluster and $B,D,F$ in the other. For simplicity we will use hard clustering where the probability of cluster membership is either 0 or 1 for each observation.  If both models have equal probability, the $\mathbf{S}$ matrix will be 
 \[
\begin{pmatrix}
1.0 & 0.5&  1.0&  0.0&  0.5&  0.0\\
  0.5 & 1.0 & 0.5 & 0.5 & 0.0 & 0.5\\
  1.0 & 0.5 & 1.0 & 0.0 & 0.5 & 0.0\\
  0.0 & 0.5 & 0.0 & 1.0 & 0.5 & 1.0\\
  0.5 & 0.0 & 0.5&  0.5 & 1.0 & 0.5\\
  0.0 & 0.5 & 0.0 & 1.0 & 0.5 & 1.0\\
\end{pmatrix}
\]
If we proceed to implement hierarchical clustering on the corresponding dissimilarity matrix, with complete linkage, we get the dendrogram shown in Figure~\ref{fig:Ch4toyHeat}. 

Averaging the two models and cutting the resulting dendogram at any 
probability level above 0.5 leads to a four-cluster solution. This makes sense.
Both clusterings agree that $A$ and $C$ belong in the same cluster as do $D$ and $F$. However, they disagree on $B$ and $E$. This dendrogram gives a probabilistic representation of this situation. If we cut the dendogram in Figure~\ref{fig:Ch4toyHeat} at 0.2, we see that $A$, $B$ and $C$ have at least a probability 0.2 of being in the same group, but cutting at 0.8 shows that $A$ and $C$ have at least a probability of 0.8 of being in the same group.

\subsection{Summary of method}
We can now carry out Bayesian model averaging by using the posterior model probabilities estimated from Equation~\ref{eq:Ch4PMP}  as weights and the similarity matrices for each candidate model, defined in Equation~\ref{eq:Ch4zzt}.  For each pair of observations $i$ and $j$ in our dataset, we propose assigning the probability vector
\begin{eqnarray}
& & \mathbb{P}\left\{\text{Observation } i,j \text{ in same cluster }|\text{ Data}\right\}\nonumber \\
& =& \sum_{m=1}^M \mathbb{P}\left\{\text{Obs } i,j \text{ in same cluster } |{\cal M}_m\right\}\mathbb{P}\left\{{\cal M}_m|\text{ Data}\right\} \nonumber\\
& \approx &\frac{ \sum_{m=1}^{M} \sum_{g=1}^{G}z_{ig}^{m}z_{jg}^{m}\exp\left(\frac{1}{2}BIC_m\right)}{\sum_{m=1}^{M}\exp\left(\frac{1}{2}BIC_m\right)} . 
\nonumber
\end{eqnarray}



\section{Results}
\label{sec:Ch4results}
The proposed methodology is demonstrated using two well known data sets: Fisher's iris data \citep{fisher1936} and Forina's wine data \citep{forina1986}. 

We clustered the datasets using the {\sf mclust} software (Version 4.4) \citep{fraley2012} and {\sf R} \citep{R2014}. Where appropriate we have ordered the observations using the {\sf gclus} R package (Version 1.3.1) \citep{hurley2012} and the {\sf seriation} R package (Version 1.0-14) \citep{hahsler2014}.

When using the default settings in {\sf mclust}, a total of 83 candidate models were fitted. These include three one-cluster models ($G=1$) and the ten possible covariance structures (Figure~\ref{fig:Ch4ellipses}) for each number of clusters $G=2,3,\ldots, 9$. One can of course fit a larger or smaller set of models if desired. 


\subsection{Clustering Fisher's iris data}
The iris data \citep{fisher1936} gives the measurements in centimetres of the variables sepal length and width and petal length and width for 150 observations with 50 of each of three species of iris: \emph{versicolor, setosa} and \emph{virginica}. 
Table~\ref{tab:Ch4iris} shows the model-based clustering models with the highest BIC values for the iris data. 
These results show that the VEV model with two clusters has the highest BIC, 
but that there is considerable uncertainty about whether this is the 
best model. In a single model scenario, the VEV model with two clusters 
would be selected and clustering would be based solely on this model.

\begin{table}[h]
\center
\caption[Iris data: BIC and posterior model probabilities for the three most favored models, i.e., those with the highest BIC. The boldfaced model is chosen by this criterion]{Iris data: BIC and posterior model probabilities for the three most favored models, i.e., those with the highest BIC. The boldfaced model is chosen by this criterion.}
\begin{tabular}{|l|l|c|c|}
\hline
Model&No of &BIC&Posterior\\
Type&Clusters&&Model Probability\\
\hline
 \bf{VEV}&\bf{2}&\bf{-561.73}&\bf{0.601}\\
VEV&3&-562.55&0.398\\
VVV&2&-574.028&0.001\\
\hline
Others&&&$<0.00001$\\
\hline
\end{tabular}
\label{tab:Ch4iris}

\end{table}

However, it can be seen from Table~\ref{tab:Ch4iris}, that the BIC value for the three-cluster VEV solution is similar to that for the selected model.
The posterior model probabilities in the table are estimated using Equation~\ref{eq:Ch4PMP} and we see that the posterior probability for this second best model is almost 40\%. 

\begin{figure}
\centering     
\subfigure[]{\label{fig:Ch4BestIrisHeat}\includegraphics[width=0.32\textwidth]{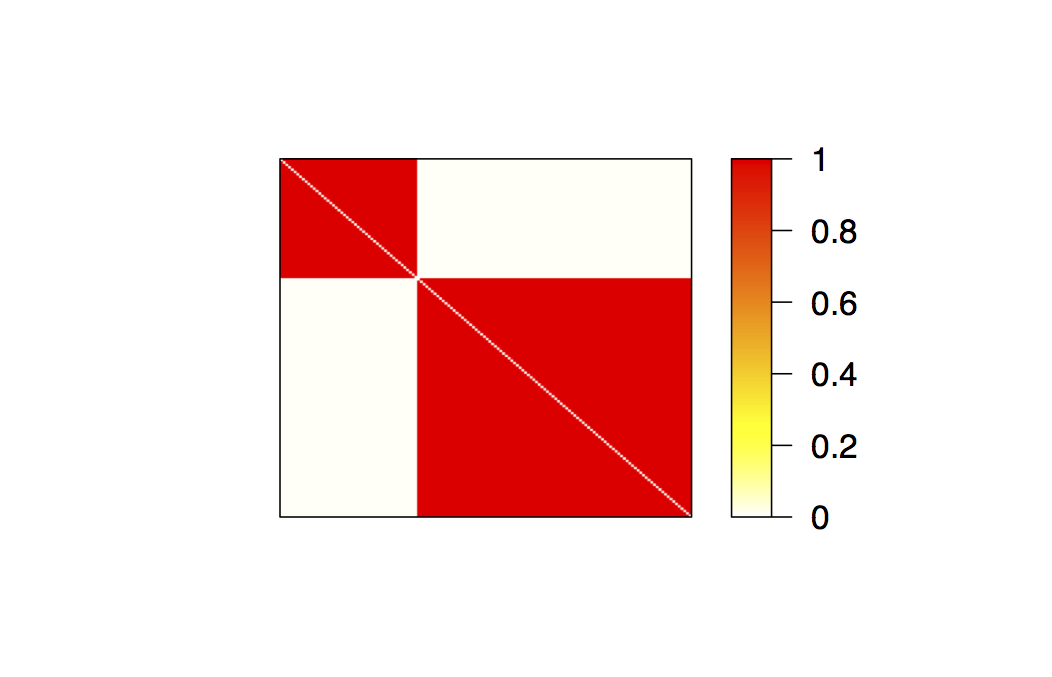}}
\subfigure[]{\label{fig:Ch42ndIrisHeat}\includegraphics[width=0.32\textwidth]{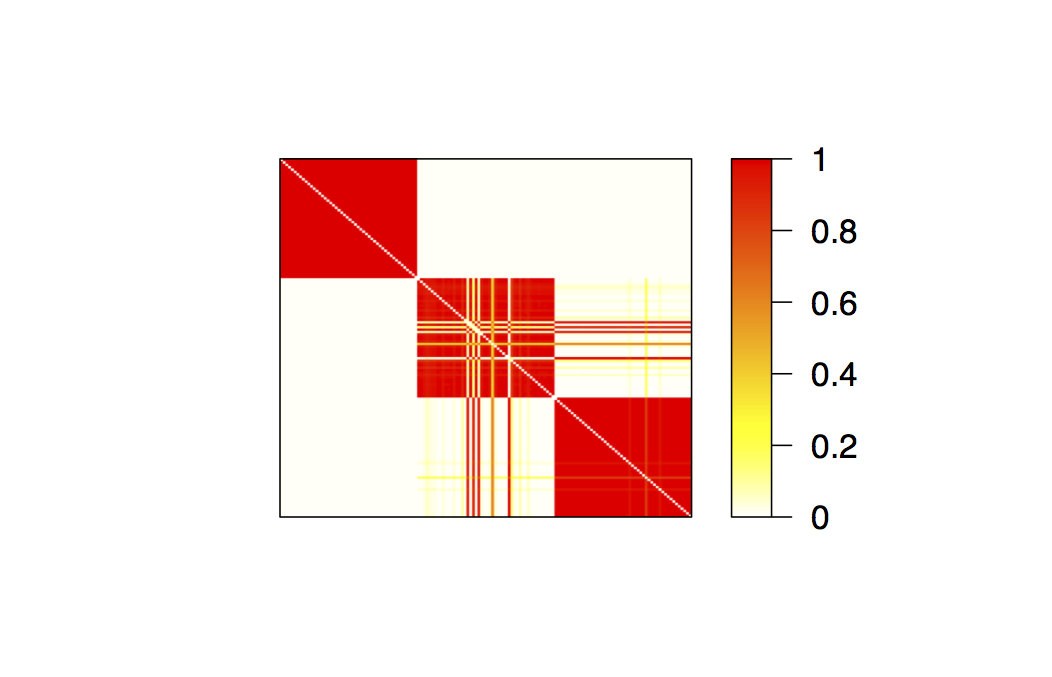}}
\subfigure[]{\label{fig:Ch4BMAIrisHeat}\includegraphics[width=0.32\textwidth]{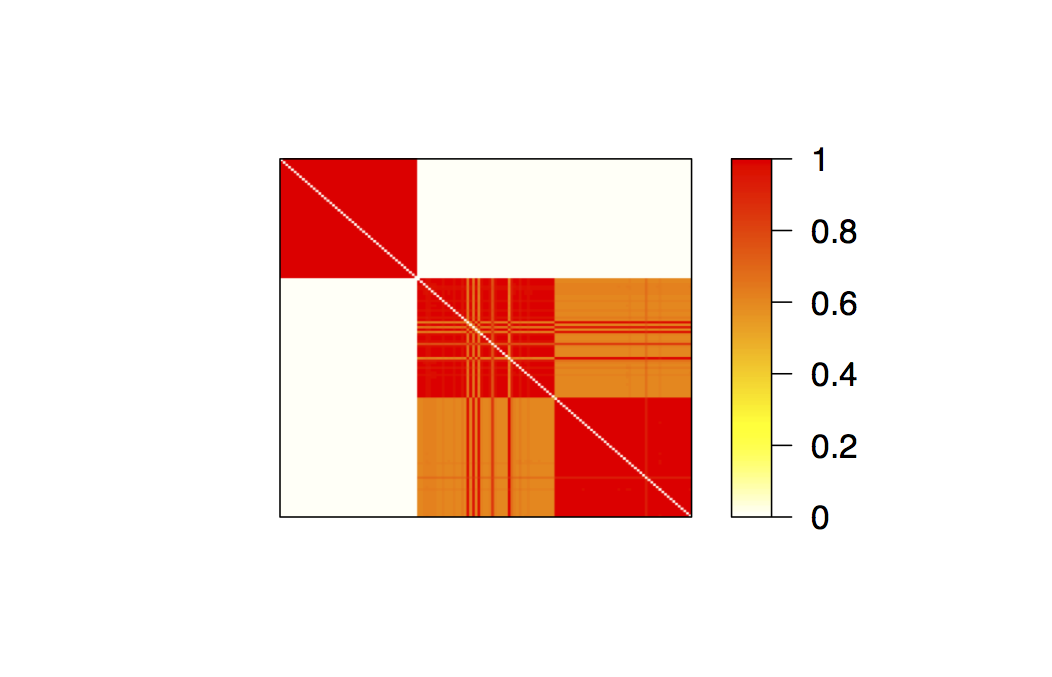}}
\caption[Iris data: Heat maps of the two models with the highest BIC, along with one for the combination of models using BMA]{Iris data: (a) Models with the highest BIC (VEV with two clusters) and (b) with the second highest BIC (VEV with three clusters) represented by heat maps of the similarity matrices.  The data are in species order. The redder the color the more likely that pairs of observations appear in the same cluster. (c) represents the combination of models using BMA.}
\label{fig:Ch4IrisHeat} 
\end{figure}

In order to visualize the observations that are likely to be in the same cluster, we use a heatmap to represent the similarity matrix. 
The ordering of the observations in the heatmap is important. 
Here we present the data in species order, which yields an intuitive heatmap, as shown in Figure~\ref{fig:Ch4IrisHeat}. However, in many applications this will not be the case, but the structure may be easier to see if the data 
are ordered using some seriation method.

Figure~\ref{fig:Ch4BestIrisHeat} shows the heatmap for the similarity matrix for the two-cluster VEV model, while Figure~\ref{fig:Ch42ndIrisHeat} shows the three-cluster model; these are the two models with the highest BIC values. The red sections of the figure represent clusters of $(i,j)$ pairs that have a high probability of being in the same cluster, which the white areas show $(i,j)$ pairs that are unlikely to be in the same cluster. The clusters are isolated along the diagonal of the heatmap. 

The model that would be selected by BIC separates the \emph{setosa} species very well from the other two species but it merges the two other species into one cluster. It can be seen also that the probabilities assigned to the co-clustering pairs are very high. The fact that there is large uncertainty associated with the two-cluster solution is not clear in the single model results. 

The second most likely model shows separation between the three clusters with high probability, as we see in Figure~\ref{fig:Ch42ndIrisHeat}. 
Thus the argument can be made that this model should contribute to the final clustering result.

Figure~\ref{fig:Ch4BMAIrisHeat} shows the heatmap for the similarity matrix produced from the Bayesian model averaging process. The method separates the large cluster into two clusters which approximate the known species groups quite well,
but also reflect the uncertainty about whether there are really three species.

\begin{figure}
\centering     
\subfigure[]{\label{fig:Ch4BestIrisDend}\includegraphics[width=0.48\textwidth]{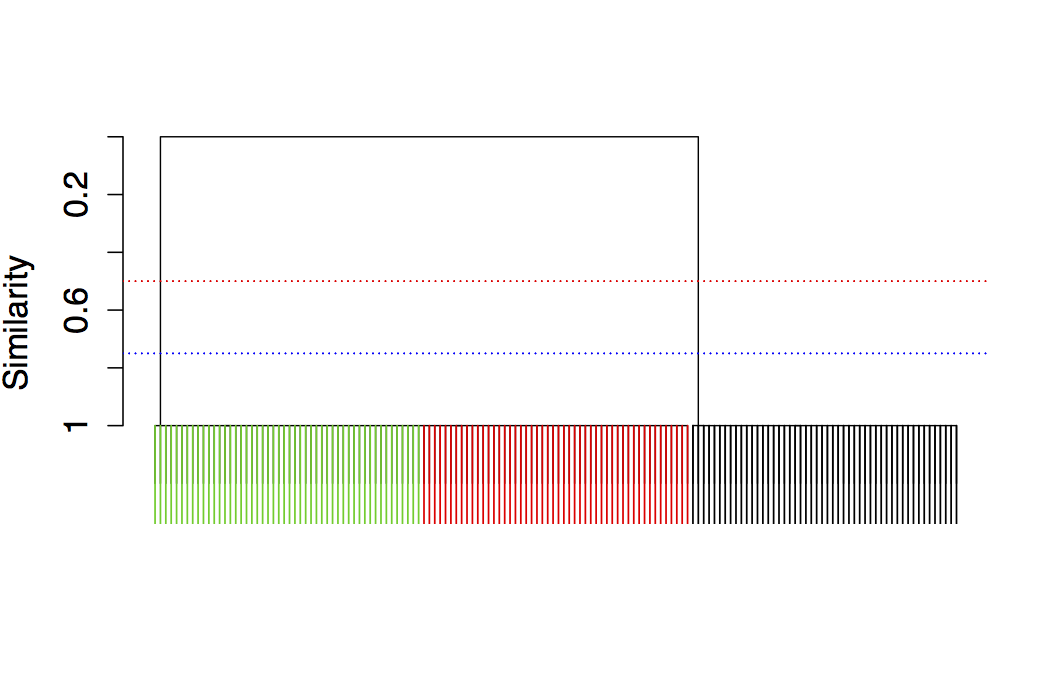}}
\subfigure[]{\label{fig:Ch4BMAIrisDend}\includegraphics[width=0.48\textwidth]{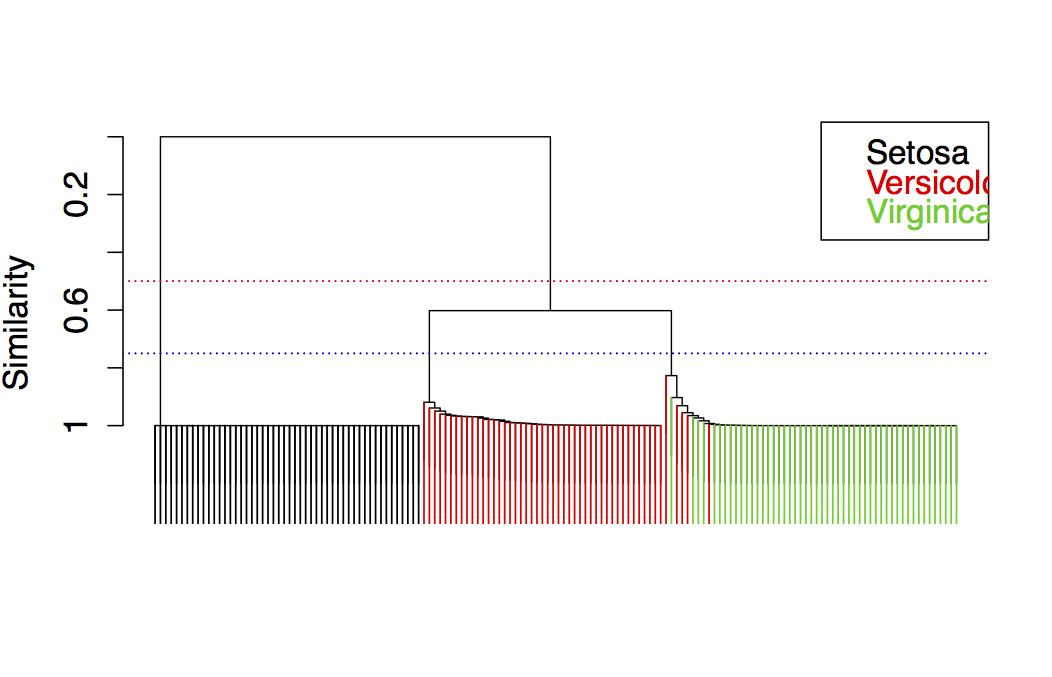}}
\caption[Iris data: Dendrograms of dissimilarity matrices for the iris data for (a) the model with the highest BIC  and (b) for the BMA solution]{Iris data: Dendrograms of dissimilarity matrices for the iris data for the model with (a) the highest BIC and (b) for the BMA solution. Cutting dendrogram (b) at a level of 0.5 (red line) and 0.75 (blue line) give a probability of at least 0.5 that the two clusters belong together, and a probability of at least 0.75 that the observations in the three branches of the tree belong in their three clusters respectively.} 
\label{fig:Ch4IrisDend}
\end{figure}

We can also present these results using dendrograms, as detailed in Section~\ref{sec:toy}, and these are shown in Figure~\ref{fig:Ch4IrisDend}. The dendrograms show that the single model results cluster the observations into two clear groups, whereas the BMA results show the possibility of both a two-cluster
and/or three-cluster solution, depending on the cutoff used.

\subsection{Clustering Italian wines data}

We now apply the methodology to a well-known dataset consisting of 27 chemical measurements on each of 178 wine samples belonging to three cultivars of wine (Barolo, Grignolino and Barbera) produced in the Piedmont region of Italy
\citep{forina1986}.
Table~\ref{tab:Ch4cultivar} shows the number of samples collected in each vintage year for the three cultivars.

\begin{table}[H]
\center
\caption[Wines data:  A list of the number of samples of each cultivar for each vintage year investigated in the study]{Wines data:  A list of the number of samples of each cultivar for each vintage year investigated in the study. }
\begin{tabular}{|c|l|rrrrrrrrrr|c|}
\hline
&&\multicolumn{10}{c}{Samples per year}&\\
Cat.index&Cat.name&'70&'71&'72&'73&'74&'75&'76&'77&'78&'79&Total\\
\hline
1&Barolo&&19&&20&20&&&&&&59\\
2&Grignolino&9&9&7&9&16&9&12&&&&71\\
3&Barbera&&&&&9&&5&&29&5&48\\
\hline
\end{tabular}
\label{tab:Ch4cultivar}
\end{table}

Table~\ref{tab:Ch4wines} shows the candidate models with the highest BIC and hence the highest approximate posterior probability; all other models had negligible posterior probability. 
Although the data consist of samples from three different cultivars, a seven-cluster model was preferred to any of the three-cluster models. 
The seven clusters successfully separate the three cultivars, and also 
partly reflect the different vintage years shown in Table \ref{tab:Ch4cultivar}
\citep{mcnicholas2008}.

\begin{table}[h]
\center
\caption[Wines data: BIC and posterior model probabilities for the three models with the highest BIC.]{Wines data: BIC and posterior model probabilities for the three models with the highest BIC. Again, the boldfaced model is chosen when carrying out model-based clustering.}
\begin{tabular}{|l|l|c|c|}
\hline
Model&No of &BIC&Posterior\\
Type&Clusters&&Model Probability\\
\hline
\bf{VEI}&7&\bf{-23951.91}&\bf{0.600}\\
EVI&3&-23953.87&0.225\\
VVI&3&-23954.37&0.175\\
\hline
others&&&$<0.001$\\
\hline
\end{tabular}
\label{tab:Ch4wines}

\end{table}

\begin{figure}
\centering     
\subfigure[]{\label{fig:Ch4BestWinesHeatns}\includegraphics[width=0.48\textwidth]{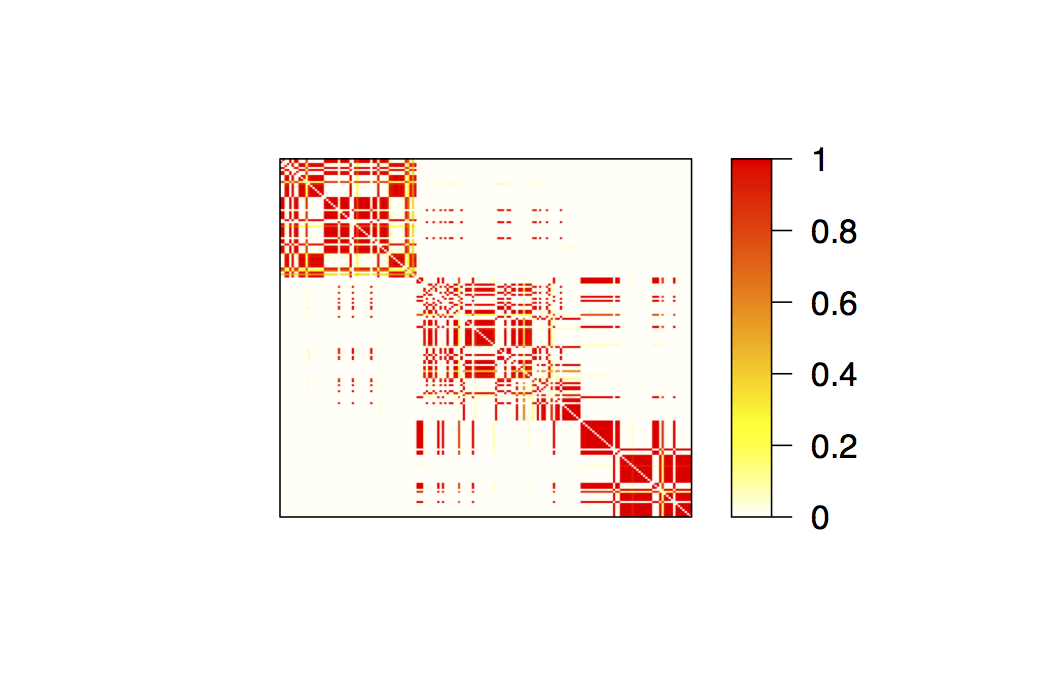}}
\subfigure[]{\label{fig:Ch4BestWinesHeatser}\includegraphics[width=0.48\textwidth]{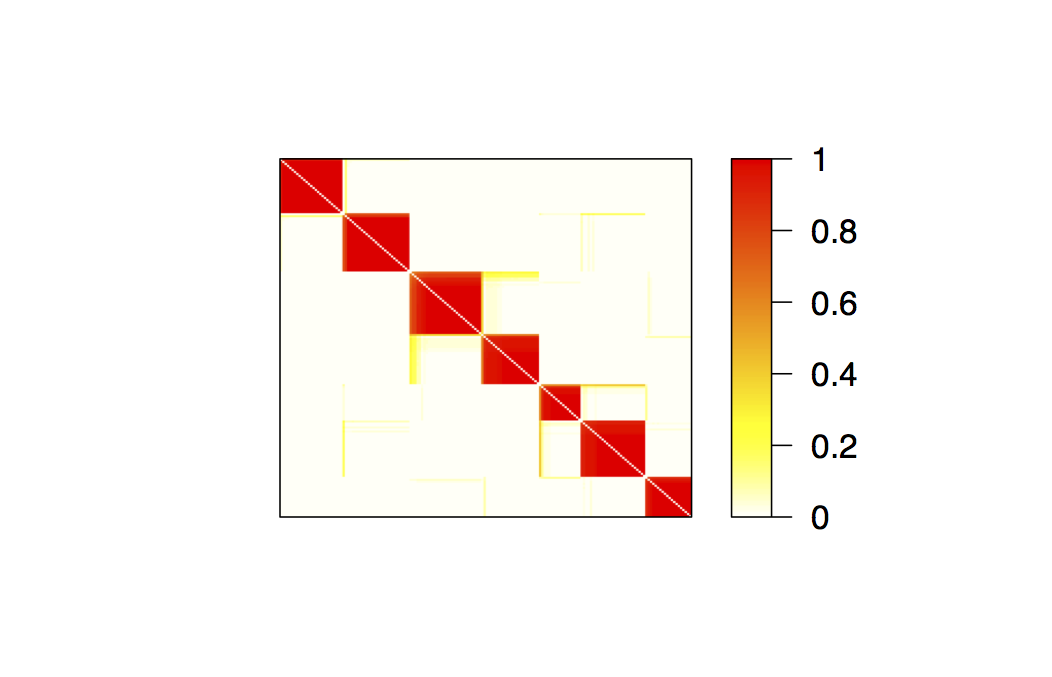}}
\caption{Wines data: (a) Highest BIC model (VEI with seven clusters) represented by heat map of the similarity matrix, in vintage year within cultivar order and (b) in seriated order.} 
\label{fig:Ch4BestWinesHeat}
\end{figure}

The heatmap of the similarity matrix of the optimal model is presented in Figure~\ref{fig:Ch4BestWinesHeatns}. Here the observations are shown in order of vintage year within cultivar. It appears that the clustering results partly
reflect the vintage years as well as the cultivars.

We seriated the data to reorder the observations, using the order of the 
leaf nodes in a dendrogram produced by hierarchical clustering with 
complete linkage. 
In hierarchical displays, a decision is needed at each merge to specify which subtree should go to the left and which to the right \citep{hurley2012}. 
We used the order suggested in \cite{gruvaeus1972}, which ensures that objects at the boundaries of each class were located next to objects outside the class which they most resembled \citep{gordon1987}.  At a merge of clusters $A$ and $B$, the new cluster is one of $(A,B)$, $(A',B)$, $(A,B')$, $(A',B')$, where $A'$ denotes $A$ in reverse order. The new cluster is chosen to minimize the distance between the object in $A$ placed adjacent to an object from $B$. 
The reordered similarity matrix, shown in Figure~\ref{fig:Ch4BestWinesHeatser},
has seven clear clusters, with uncertainty about the group membership of
some of the observations.

\begin{figure}[h]
\centering     
\subfigure[]{\label{fig:Ch4BMAWinesHeatns}\includegraphics[width=0.48\textwidth]{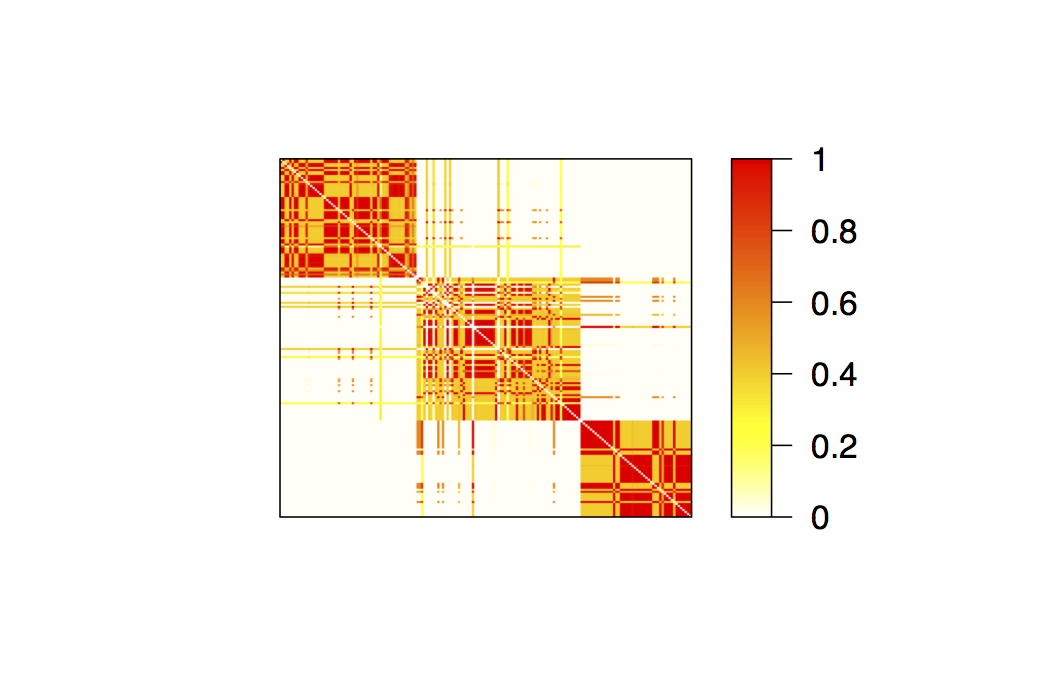}}
\subfigure[]{\label{fig:Ch4BMAWinesHeatser}\includegraphics[width=0.48\textwidth]{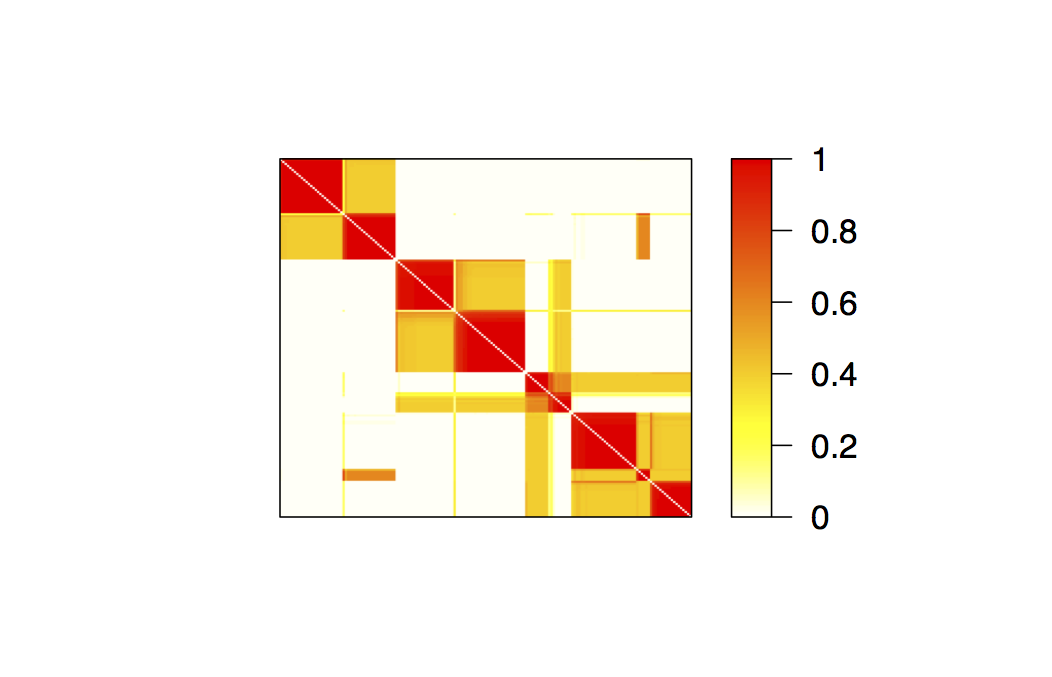}}
\caption[Wines data: Heat maps after Bayesian model averaging denoting the combined similarity matrix (consensus matrix) (a) in vintage year within cultivar order and (b) in seriated order.]{Wines data: Heat maps after Bayesian model averaging denoting the combined similarity matrix (consensus matrix) (a) in vintage year within cultivar order and (b) in seriated order.}
\label{fig:Ch4BMAWinesHeat}
\end{figure}

From Table~\ref{tab:Ch4wines} we see that the combined posterior probability 
of the two three-cluster models (EVI and VVI) is high, at just under 40\%. 
We use BMA to average the similarity matrices across the candidate models 
and show the results in Figure~\ref{fig:Ch4BMAWinesHeat}.
Figure~\ref{fig:Ch4BMAWinesHeatns} shows the clustering in the originally presented order, after BMA has been carried out. 
It gives a clearer picture of the cultivar clustering, although there is some uncertainty, shown by the yellow colors.  
Figure~\ref{fig:Ch4BMAWinesHeatser} shows the seriated version, where the smaller clusters group naturally into three larger clusters, but with one small 
cluster that could belong to either of two cultivars.

\begin{figure}
\centering     
\subfigure[]{\label{fig:Ch4BestWinesDend}\includegraphics[width=0.48\textwidth]{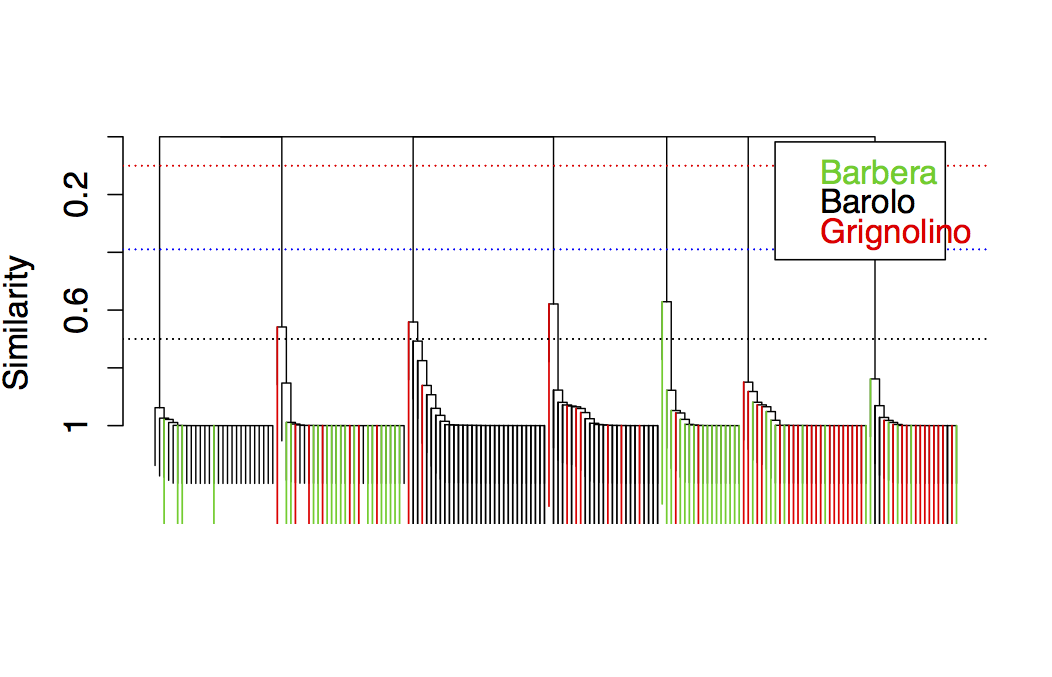}}
\subfigure[]{\label{fig:Ch4BMAWinesDend}\includegraphics[width=0.48\textwidth]{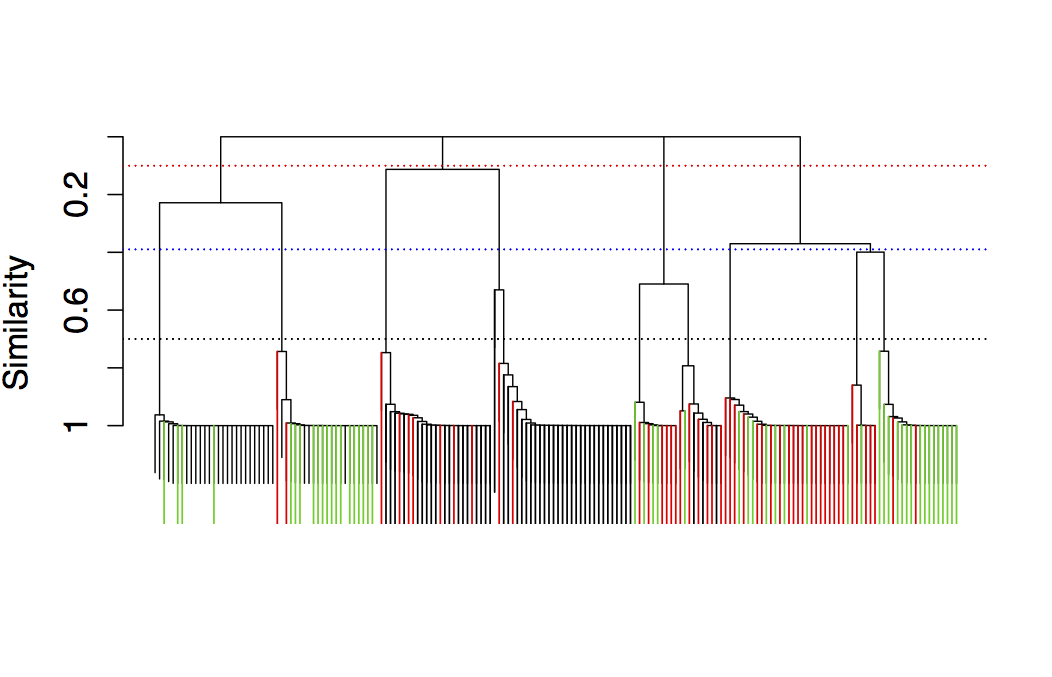}}
\caption[Wines data: Dendrograms of dissimilarity matrices for the wines data for the model with (a) the highest BIC (VEI with seven clusters) and (b) for the BMA solution.]{Wines data: Dendrograms of dissimilarity matrices for (a) the model with the highest BIC (VEI with seven clusters), and (b) the BMA solution. 
The horizontal dashed lines indicate the partitions that arise by 
cutting the dendogram at various levels.
The Grignolino cultivar is the red one on the right.}
\label{fig:Ch4WinesDend}
\end{figure}

Figure~\ref{fig:Ch4WinesDend} shows the dendrogram of the complete-linkage clustering of the consensus matrices according to (a) the best model and (b) BMA.
The dendogram for the best model (VEI with seven clusters) reproduces
the model's seven clusters quite clearly if the dendogram is cut at any level
above about 0.5, as expected.
The BMA dendogram reflects more uncertainty, as one would expect.
It divides the data into four groups if cut at 0.1.

The two three-cluster solutions give similar clustering results.
In the seven-cluster solution, three of the clusters are for the Grignolino observations, and two each for the Barbera and  Barolo observations. The Barolo and Barbera clusters break down by vintage year, whereas for the Grignolino observations the clusters correspond less clearly to vintage years. 

\section{Bayesian model averaging for density estimation.}
\label{sec:Ch4density}
The fitted mixture model produced by model-based clustering provides an overall density estimate for the data generation process. \citet{ferguson1983} showed that finite mixtures of normal distributions can approximate any distribution on the real line to within any given accuracy.  Thus the density estimate produced by model-based clustering can be used as a density estimation method.
The performance of densities fitted in this way was assessed by \citet{roeder1997} for the univariate case and by \citet{fraley2002} for the bivariate case. 
Both of these studies showed model-based clustering to be competitive with state of the art kernel density estimation methods.

We propose using Bayesian model averaging of the density estimates for each model using the posterior model probabilities from Equation~\ref{eq:Ch4PMP}. 
So, given model-based density estimates $\hat{f}_{{\cal M}_m}(x)$, for $m=1,2,\ldots, M$, we have
\begin{equation*}
\hat{f}_{\text{BMA}}(x)=\sum_{m=1}^{M} \mathbb{P}\{{\cal M}_m|\text{ Data }\} \hat{f}_{{\cal M}_m}(x).
\end{equation*}

We compare the BMA results with kernel density estimation methods.
Kernel density estimation has long been used for density estimation and visualization of univariate data \citep{rosenblatt1956,parzen1962,jones1996}. 
The kernel density estimate $\hat{f}_h$ of a univariate density $f$ based on a random sample $X_1,\dots, X_N$ of size $N$ is 
\begin{eqnarray}
\hat{f}_h(x)&=&\frac{1}{N} \sum_{i=1}^{N} K_h(x-X_i) \nonumber\\
&=&\frac{1}{N} \sum_{i=1}^{N} \frac{1}{h}K\left(\frac{x-X_i}{h}\right) ,
\nonumber
\end{eqnarray}
where $K_h(\cdot)= (1/h)K(\cdot/h)$ for a kernel function $K$, assumed to be a symmetric probability density and a bandwidth $h$ (the smoothing parameter).
We also compare $\hat{f}_{\text{BMA}}$ with the single-model estimate
$\hat{f}_{\text{SM}}$, where
\begin{equation*}
\hat{f}_{\text{SM}}(x)=\hat{f}_{\widehat{\cal M}_m}(x)
\end{equation*}
where $\widehat{\cal M}_m$ is the model with the highest value of BIC.
(SM stands for ``single model''.)  

\citet{wand1994} described the extension of the method to the bivariate case.
The multivariate kernel density estimate is 
\begin{equation*}
\hat{f}(x; \mathbf{H})=\frac{1}{N} \sum_{i=1}^{N} K_{\mathbf{H}}(x-X_i),
\end{equation*}
where $\mathbf{H}$ is now a $d \times d$ positive definite matrix and $K$ is a $d$-variate spherically symmetric density function; a common simplification 
is to use a diagonal $\mathbf{H}$ \citep[e.g.,][Chapter 4.2]{wand1994}.

The performance of density estimation procedures can be compared using mean integrated squared error (MISE) or expected Kullback-Leibler divergence (KL),
defined as follows:
\begin{eqnarray}
\mbox{MISE} &=& {\mathbb E}^{\hat{f}}\int [f(x)-\hat{f}(x)]^2dx
\label{eq:MISE} \\
\mbox{MKL} &=& {\mathbb E}^{\hat{f}}\int \log\left[\frac{f(x)}{\hat{f}(x)}\right]f(x)dx,
\label{eq:KL}
\end{eqnarray}
where the expected value is taken with respect to the resulting density 
estimate, $\hat{f}$, computed from a random sample, $X_1,X_2,\ldots,X_n$, 
drawn from $f$.  For both criteria, smaller is better.
MISE (Equation~\ref{eq:MISE}) is the most commonly used measure of performance, while Kullback-Leibler divergence (Equation~\ref{eq:KL}) provides an alternative which considers the differences in densities on a logarithmic scale; this places more emphasis on differences in regions of low density.

The asymptotic performance of kernel density estimation under the MISE criterion was described by \cite{silverman1986} and \cite{scott1992} and the form of the optimal bandwidth for the kernel was derived. \cite{hall1987} studied the asymptotic performance of kernel density estimation under Kullback-Leibler divergence and showed that the optimal bandwidth in this case leads to a smoother density estimate than under MISE. The optimal bandwidths differ because Kullback-Leibler puts a larger penalty on regions where a density estimate has low density but the true density has high density. 
\citet{wand1993} and \citet{duong2003} investigated the use of unconstrained parametrisations of the $\mathbf{H}$ matrix as opposed to diagonal ones on simulated target densities, and concluded that this can improve efficiency.

We use the extended version of multivariate kernel density estimation as described in \cite{duong2003} and \cite{duong2007} as a benchmark for comparison with the model-based clustering density estimates. This approach uses a two-stage estimation procedure, where a pilot density estimate is used to get an improved estimate of the optimal bandwidth compared to the standard plug-in estimates \cite[e.g.][Section 3.4.2]{silverman1986}. \cite{duong2007} used kernels with non-diagonal ${\mathbf H}$ matrices.

\subsection{Density estimation results}
\subsubsection{Simulation study for bivariate density estimation}
\label{sec:bivarsim}
\cite{fraley2002} defined bivariate analogs of the first ten of the univariate datasets defined in \citet{marron1992} to compare the performance of the fitted model-based clustering density with multivariate kernel density estimation. Contour plots of the densities are shown in Appendix~\ref{sec:settings} as well as the parameter settings for the actual densities we used.

We produced 250 simulations of 250 observations from each of the same ten distributions  and compared density estimation using model-based clustering with the Bayesian model averaged result in terms of the mean integrated squared error and the Kullback-Leibler distance. 
We compared the performance of the model-based clustering approaches to the extended kernel density estimaton method described in \cite{duong2003} as implemented in the {\sf ks} R package \citep{duong2007,duong2014}. 

The results of the simulation study are shown in Table ~\ref{tab:DEresults} where the numbers in the MISE columns are the MISE for kernel density estimation and for the single model respectively divided by the MISE for the BMA method. 
Similarly, for the Kullback-Leibler column, we divide the Kullback-Leibler distance for kernel density estimation and for the single model by the Kullback-Leibler distance for the BMA method. 

\begin{table}[H]
\center
\caption[Comparison of multivariate density estimation methods]{Mean MISE and Kullback-Leibler (KL) distance ratios for density estimation via model-based clustering with Bayesian model averaging (BMA) as against kernel density estimation with the {\sf ks} R package (KS) and single-model model-based clustering (SM). Datasets used are the ten bivariate extensions of Marron-Wand univariate densities from \cite{fraley2002}. 
The numbers in the MISE columns are the MISE for kernel density estimation and for the single model respectively divided by the MISE for the BMA method;
similarly for the KL columns.  
A value greater than 1.00 indicates that BMA is preferred to the competing method while a value less than 1.00 denotes that the competing method is preferred
to BMA.  (Results are based on 250 simulated datasets with 
250 observations each for each density type.) }
\begin{tabular}{|l|cc|cc|}
\hline
Model&\multicolumn{2}{c|}{KS/BMA}&\multicolumn{2}{c|}{SM/BMA}\\
&MISE&KL&MISE&KL\\
\hline
Single Gaussian&6.19&4.86&1.00&1.00\\
Skewed unimodal&1.06&1.53&1.03&1.03\\
Strongly skewed&2.75&11.4&1.04&1.05\\
Kurtotic unimodal& 0.45&11.7&1.00&1.00\\
Outlier&1.81&1.68&1.00&1.00\\
Bimodal&1.12&1.46&1.08&1.08\\
Separated bimodal&3.10&3.74&1.01&1.01\\
Asymmetric bimodal&0.55&2.63&1.00&1.00\\
Trimodal&0.98&0.99&1.01&1.03\\
Claw (Bart Simpson) & 0.91&1.37&1.09&1.19\\
\hline
\end{tabular}
\label{tab:DEresults}
\end{table}

Density estimation using model-based clustering with BMA compares
very favorably with kernel density estimation according to the KL criterion,
while it compares favorably in the majority of cases by the MISE criterion.
It also compares favorably with density estimation using model-based 
clustering with a single model in all cases and by both criteria,
although the gains are more modest.

\cite{fraley2002} gave further comparisons of single model density estimation with other kernel density estimation methods, namely Gaussian kernel density estimation using both the normal optimal bandwidth and cross-validated bandwidth \citep{bowman1997}.


\subsubsection{Simulation study for higher dimensional density estimation}
We also investigated density estimation for higher dimensional data. We first added dimensions consisting of observations from the standard normal distribution to the existing distributions used in Section~\ref{sec:bivarsim}. The first three rows in Table~\ref{tab:HDEresults} refer to bivariate distributions with one extra such dimension added, while the next three rows refer to the same distributions with four extra standard normal dimensions. 
We also implemented three- and six-dimensional versions of the bimodal distribution. We allowed for variable separation of the modes as in the bimodal case, and the results are in rows seven to twelve of Table~~\ref{tab:HDEresults}. The settings used for all these simulations are in Appendix~\ref{sec:highdimensional}.

\begin{table}[H]
\center
\caption[Comparison of density estimation methods using three- and six-dimensional distributions.]{MISE and Kullback-Leibler (KL) distance ratios for density estimation via model-based clustering with Bayesian model averaging (BMA) as against kernel density estimation with the {\sf ks}  package (KS) and single model model-based clustering (SM). The top six lines comprise simulations of bivariate data as before with either one or four extra dimensions of standard normal. The bottom six lines comprise extensions to three and six dimensions of the bimodal density of \protect\cite{fraley2002} with variable separation of the modes.}
\begin{tabular}{|l|cc|cc|}
\hline
Model&\multicolumn{2}{c|}{KS/BMA}&\multicolumn{2}{c|}{SM/BMA}\\
&MISE&KL&MISE&KL\\
\hline
Strongly skewed 3D&2.46&7.47&1.03&1.04\\
Separated bimodal 3D&5.08&6.39&1.00&1.02\\
Asymmetric bimodal 3D&0.75&3.53&1.01&1.00\\
\hline
Strongly skewed 6D&2.66&4.95&1.03&1.04\\
Separated bimodal 6D&8.18&11.0&1.03&1.03\\
Asymmetric bimodal 6D&1.41&4.31&1.06&1.04\\
\hline
\hline
Bimodal 3D (sep of 1.5)&6.51&6.45&1.00&1.03\\
Bimodal 3D (sep of 3)&1.94&3.13&1.03&1.05\\
Bimodal 3D (sep of 5)&5.34&6.30&1.00&1.00\\
\hline
Bimodal 6D (sep of 1.5)&12.3&10.56&1.06&1.04\\
Bimodal 6D (sep of 3)&10.8&13.61&1.03&1.04\\
Bimodal 6D (sep of 5)&8.41&11.9&1.11&1.08\\
\hline
\end{tabular}
\label{tab:HDEresults}
\end{table}

Model-based clustering with BMA strongly outperformed kernel density
estimation in all cases except one for both three-dimensional and 
six-dimensional data, according to both criteria, MISE and KL.
Model-based clustering with BMA also uniformly outperformed single-model
model-based clustering, although the gain was more modest.

Density estimation using model-based clustering
can be carried out for higher dimensions, and performs well.
However, we limited ourselves to six dimensions because the {\tt ks} R
package provides results for at most six dimensions.
We conjecture that model-based clustering's gain in performance 
would be even greater for higher dimensions.

\subsubsection{Lansing maples}
We now consider the Lansing trees data set \citep{gerrard1969} where we are interested in the density of the maple trees in a  forest. The models with the highest BIC are shown in Table~\ref{tab:imaples} and we can see that three models have non-negligible posterior probability. 

\begin{table}[H]
\center
\caption{Lansing Woods data: BIC and posterior model probabilities for the three models with the highest BIC.}
\begin{tabular}{|l|l|c|c|}
\hline
Model&No of &BIC&Posterior\\
Type&Clusters&&Model Probability\\
\hline
{\bf VII}&{\bf 7}&{\bf 154.339}&{\bf 0.49462}\\
VEI&7&153.569&0.33655\\
VII&6&152.081&0.15998\\
\hline
others&&&$<0.01$\\
\hline
\end{tabular}
\label{tab:imaples}
\end{table}

A six-cluster solution has approximately 16\% posterior model probability with two seven-cluster solutions having approximately 83\% between them. There are very small probabilities associated with five and eight cluster solutions respectively. However, the best model that would be chosen according to BIC is VII with seven clusters.

\begin{figure}
\centering     
\subfigure[Model VII with 7 clusters]{\label{fig:densVII7}\includegraphics[width=0.48\textwidth]{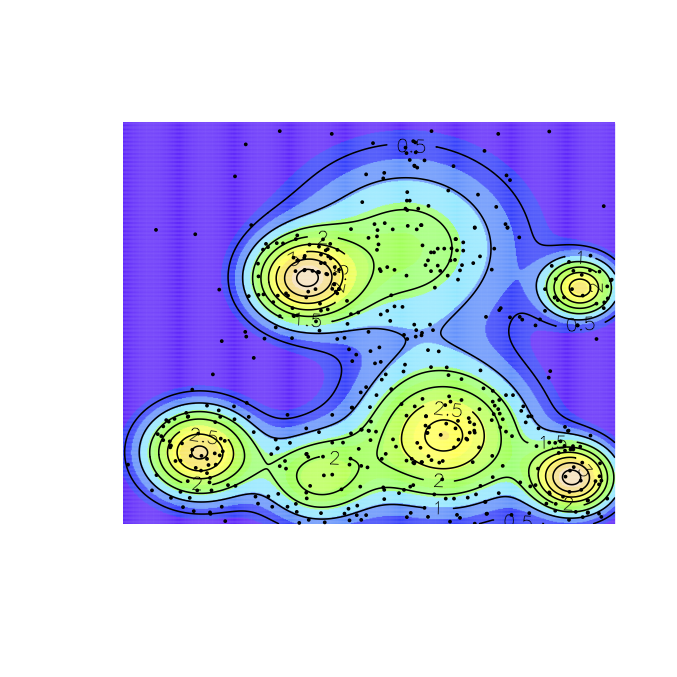}}
\subfigure[Model VEI with 7 clusters]{\label{fig:densVEI7}\includegraphics[width=0.48\textwidth]{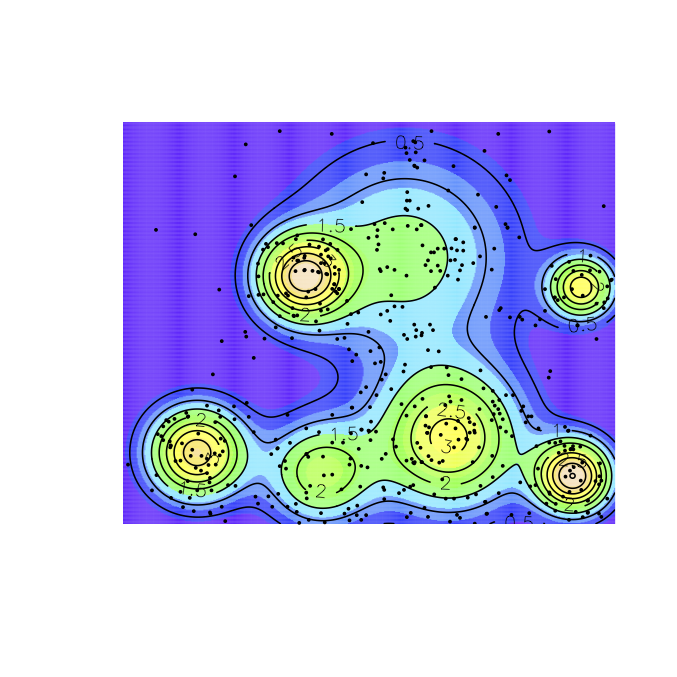}}

\subfigure[Model VII with 6 clusters]{\label{fig:densVII6}\includegraphics[width=0.48\textwidth]{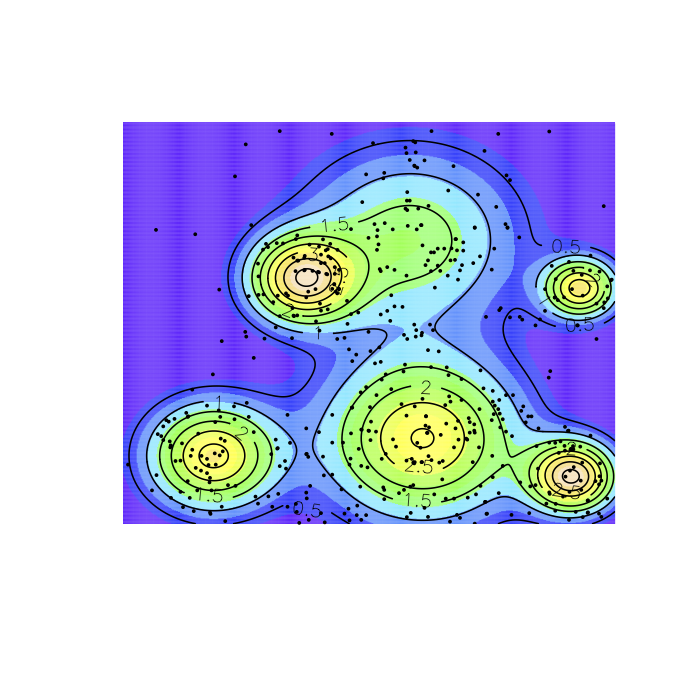}}
\subfigure[BMA]{\label{fig:densBMA}\includegraphics[width=0.48\textwidth]{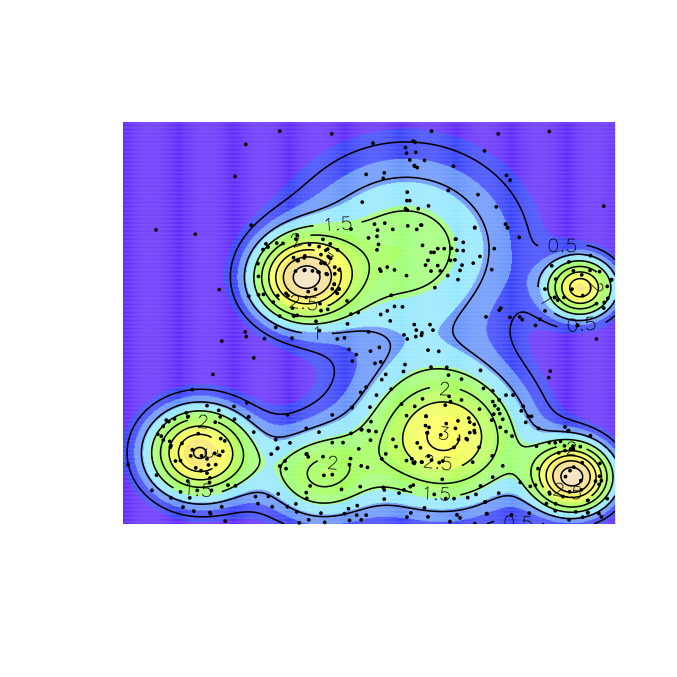}}
\caption[Lansing Woods maples: Contour plots denoting the density estimates with the three highest model probabilities (a -- c) and the Bayesian model averaged density estimate (d).]{Lansing Woods data: (a--c) Contour plots denoting the density estimates with the three highest model probabilities and (d) the Bayesian model averaged density estimate. The locations of the maple trees are overlaid.} 
\label{fig:lansing}
\end{figure}

We can compare graphically the contour plots for the models with the highest BIC and the plot for the density estmation after BIC (Figure~\ref{fig:lansing}). Figures~\ref{fig:densVII7}, \ref{fig:densVEI7} and \ref{fig:densVII6} show the three most likely models according to BIC, while Figure~\ref{fig:densBMA} shows the BMA density estimate.
We can see that the high density region at approximately $(0.4,0.1)$ in Figures~\ref{fig:densVII7} and \ref{fig:densVII7} does not appear in Figure~\ref{fig:densVII6}. Further, the density of this region is smaller in Figure~\ref{fig:densBMA}. Thus the BMA density estimate takes into account the possibility that the model density plotted in Figure~\ref{fig:densVII6} is appropriate.

\begin{figure}
\centering
\includegraphics[width=0.8\textwidth]{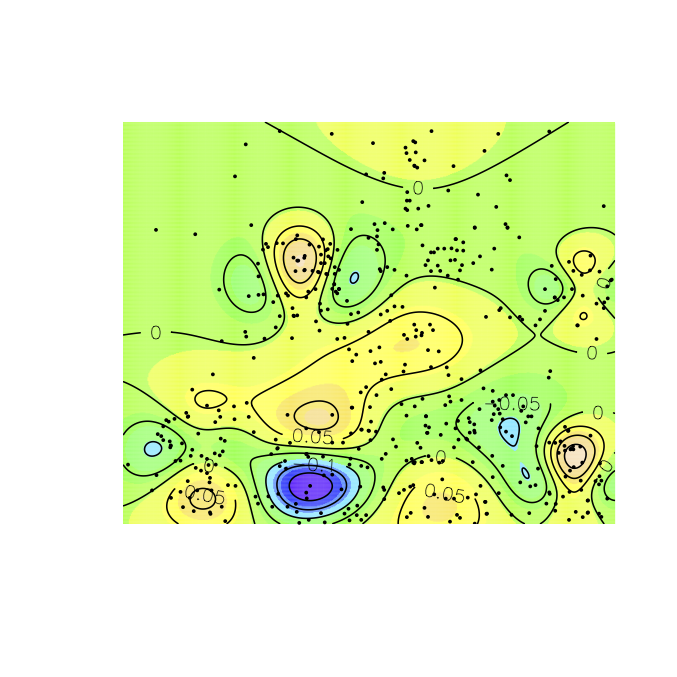}
\caption[Lansing Woods maples: Difference between the densities for the model with highest probability and after BMA]{Lansing Wood data: Difference between the densities for the model with highest probability and after BMA.}  
\label{fig:densdiff}
\end{figure}

Figure~\ref{fig:densdiff} shows the differences between the density estimates produced by BMA and the density estimate for the VII model with seven clusters.
Again, the difference in probability density around the cluster at $(0.4,0.1)$ is evident. There are some other small areas of smoothing denoted in blue. 
The brown peaks are higher densities to compensate for the lower density at $(0.4,0.1)$.

\section{Discussion}
\label{sec:Ch4discuss}
%
We have implemented Bayesian model averaging for model-based clustering and density estimation. The results show that the BMA assesses the model uncertainty better than the single best model approach.

We have proposed similarity matrices as a way of representing ensembles of clustering solutions. They provide an intuitive way of visualising clustering solutions and are consistent across models with different numbers of clusters. They are also invariant to label switching between models. Similar ideas were put forward by \cite{strehl2003} and \cite{monti2003} using binary classification matrices, while \cite{fern2003} extended the idea to soft classifications when proposing their random projection clustering method. All of these papers use different ways of combining the matrices than BMA.

\cite{kuncheva2004} suggested that ``cutting'' the matrix at a certain threshold is equivalent to running the single link algorithm and cutting the dendrogram at that level. We argue that using complete linkage gives a more intuitive interpretation.


We have shown how to apply Bayesian model averaging to clustering solutions using the similarity matrix. This provides a statistical postprocessing method which takes account of model uncertainty. When used on datasets with well-documented structure or a known ground truth, results achieved with Bayesian model averaging are consistent with the ground truth. We have shown that carrying out BMA on datasets that have no ground truth might give additional information than using model-based clustering alone and at little computational cost.


The model-based density estimation method described in \cite{fraley2002} gives better results in many cases than those given by well-known kernel density estimation methods for the simulated datasets analysed here. With the addition of the Bayesian model averaging described in this paper, even more improvement can be seen, and this becomes more pronounced in higher dimensions. 

In the case where the derivative of the density estimate is of interest (e.g. \citet{jones1994} and others), a separate bandwidth is needed for each order derivative \citep{chacon2013}. An advantage of the finite mixture density estimate and the BMA mixture density estimate is that a single estimate is produced which can be used for estimating derivatives of any order.

The proposed methodology could also be used when more than one family of component distributions is under consideration. For example, the normal mixture models with eigendecomposed covariances used in {\sf mclust} \citep{fraley2002}, the normal mixtures with factor analytic covariance structure used in {\sf pgmm} \citep{mcnicholas2008}, the multivariate $t$ mixtures used in {\sf EMMIX} \citep{mclachlan99} and the skew-$t$ mixtures used in {\sf EMMIXuskew} \citep{lee2013b} could be combined using the BMA procedure.

One previous approach to Bayesian model averaging within model-based clustering \citep{wei2014} considers noninvariance of model-based clustering results
to cluster labeling by matching the clusterings for competing models using an cluster agreement criterion. Further, they combine the clusters in the larger model so that the models being combined have the same number of clusters, $G$. 
We have proposed a very different approach for Bayesian model averaging for model-based clustering by choosing a quantity of interest, $\Delta$, that is invariant to cluster labeling and that has a common meaning for all values of $G$. 

Overall, our results suggest that Bayesian model averaging is a useful postprocessing tool for model-based clustering and density estimation. It often helps and seldom disimproves the results, so it could be used routinely as part of model-based clustering. 

\section*{Acknowledgements}
%
This research is supported by the Programme for Research In Third Level
Institutions (PRTLI) Cycle 5 and co-funded by the European Regional
Development Fund, the Science Foundation Ireland Research Walton Fellowship (11/W.1/I2079), the Science Foundation Ireland funded Insight Research Center (SFI/12/RC/2289), and NIH grants R01-HD054511, R01-HD070936 and U54-HL127624.
%

\bibliographystyle{ims}
\bibliography{refs}

\pagebreak
\appendix
\section{Settings used for density simulations}
\label{sec:settings}
\normalsize
\subsection{Bivariate extensions of the Marron and Wand distributions as used in Fraley \emph{et al. }(2002)}
\subsubsection{Single Gaussian}

￼\begin{figure}[H]
\centering
\includegraphics[width=0.7\textwidth]{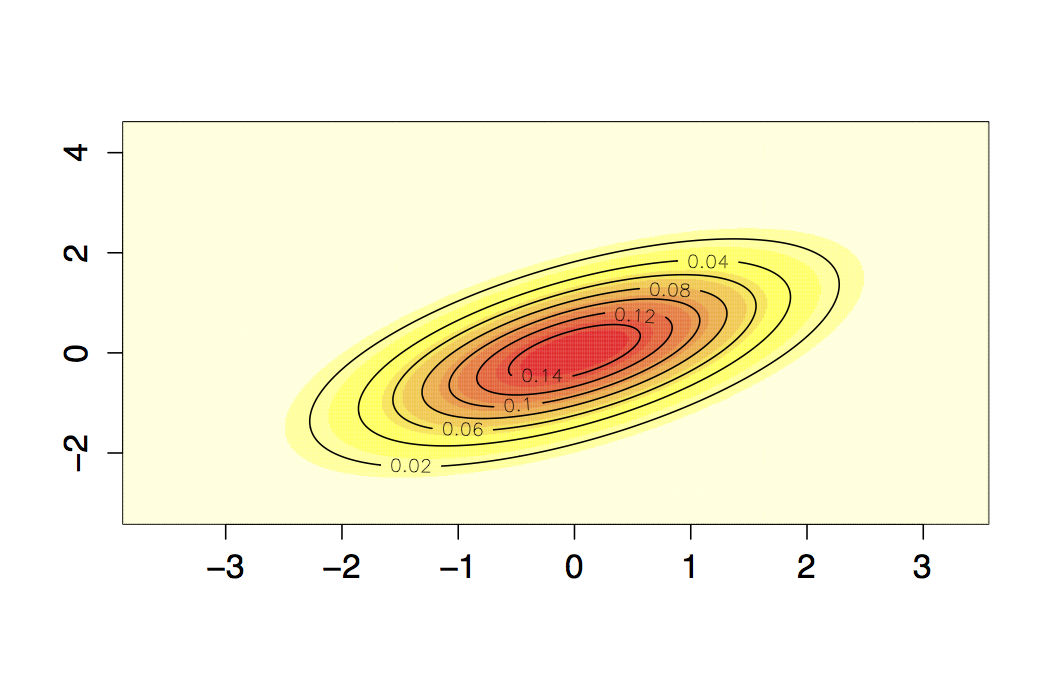}
\caption[Gaussian distribution: Contour map]{Gaussian distribution: Contour map denoting the density of a unimodal Gaussian distribution.}  
\end{figure}

\begin{table}[h]
\caption{Gaussian distribution: Simulation settings.}
\center
\begin{tabular}{|l|c|c|c|}
\hline
Cluster&Prop&Mean&Covariance\\
&$\tau_g$&$(\mu_g)$&$(\Sigma_g)$\\ 
\hline
1&1&$\begin{pmatrix}0 \\ 0 \\ \end{pmatrix}$&
$\begin{pmatrix}1.25&0.75\\0.75&1.25\\ \end{pmatrix}$\\
\hline
\end{tabular}
\label{tab:Gaussian}

\end{table}
\pagebreak
\subsubsection{Skewed unimodal}

￼\begin{figure}[H]
\centering
\includegraphics[width=0.7\textwidth]{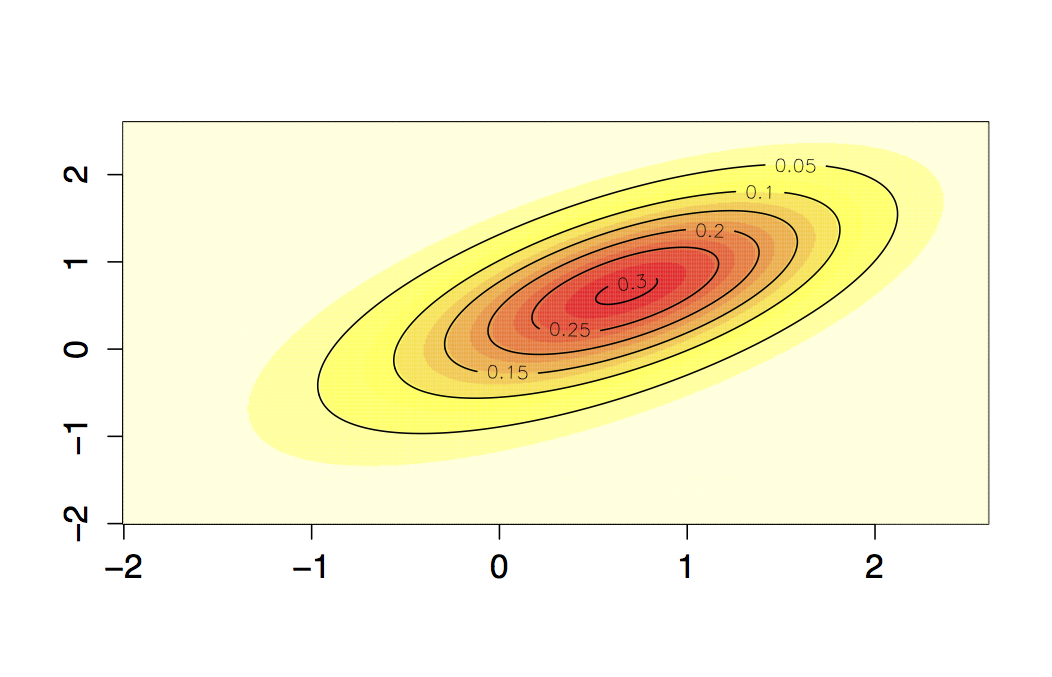}
\caption[Skewed unimodal distribution: Contour map]{Skewed unimodal distribution: Contour map denoting the density of a skewed unimodal  distribution.}  
\end{figure}

\begin{table}[h]
\caption{Skewed unimodal distribution: Simulation settings.}
\center
\begin{tabular}{|l|c|c|c|}
\hline
Cluster&Prop&Mean&Covariance\\
&$\tau_g$&$(\mu_g)$&$(\Sigma_g)$\\ 
\hline
1&1/5&$\begin{pmatrix}0 \\ 0 \\ \end{pmatrix}$&
$\begin{pmatrix}1.25&0.75\\0.75&1.25\\ \end{pmatrix}$\\
\hline
2&1/5&$\begin{pmatrix}0.3535534 \\ 0.3535534 \\ \end{pmatrix}$&
$\begin{pmatrix}0.6804138&0.4082483\\0.4082483&0.6804138\\ \end{pmatrix}$\\
\hline
3&3/5&$\begin{pmatrix}0.7660323 \\ 0.7660323 \\ \end{pmatrix}$&
$\begin{pmatrix}0.5176083&0.3105650\\0.3105650&0.5176083\\ \end{pmatrix}$\\
\hline
\end{tabular}
\label{tab:su}
\end{table}
\pagebreak
\subsubsection{Strongly skewed}

￼\begin{figure}[H]
\centering
\includegraphics[width=0.5\textwidth]{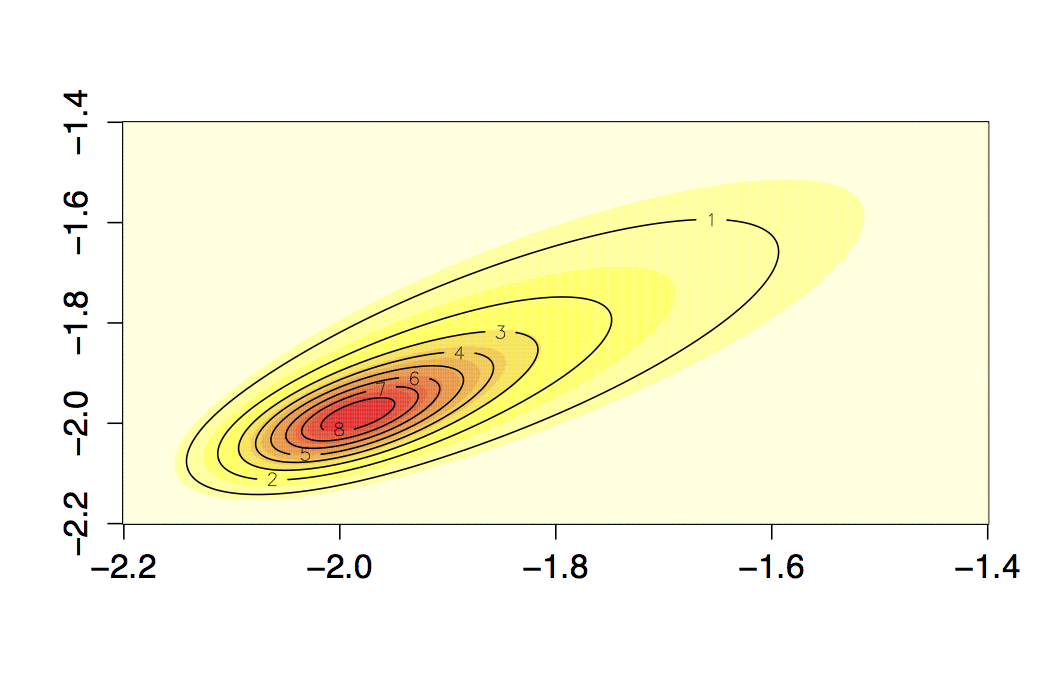}
\caption[Strongly skewed distribution: Contour map]{Strongly skewed distribution: Contour map denoting the density of a strongly skewed distribution.}  
\end{figure}

\begin{table}[h]
\caption{Strongly skewed distribution: Simulation settings.}
\center
\begin{tabular}{|l|c|c|c|}
\hline
Cluster&Prop&Mean&Covariance\\
&$\tau_g$&$(\mu_g)$&$(\Sigma_g)$\\ 
\hline
1&1/8&$\begin{pmatrix}0 \\ 0 \\ \end{pmatrix}$&
$\begin{pmatrix}1.25&0.75\\0.75&1.25\\ \end{pmatrix}$\\
\hline
2&1/8&$\begin{pmatrix}-0.7071068 \\ -0.7071068 \\ \end{pmatrix}$&
$\begin{pmatrix}0.5555556&0.3333333\\0.3333333&0.5555556\\ \end{pmatrix}$\\
\hline
3&1/8&$\begin{pmatrix}-1.178511 \\ -1.178511 \\ \end{pmatrix}$&
$\begin{pmatrix}0.2469136&0.1481481\\0.1481481&0.2469136\\ \end{pmatrix}$\\
\hline
4&1/8&$\begin{pmatrix}-1.492781 \\ -1.492781 \\ \end{pmatrix}$&
$\begin{pmatrix}0.10973937&0.06584362\\0.06584362&0.10973937\\ \end{pmatrix}$\\
\hline
5&1/8&$\begin{pmatrix}-1.702294 \\ -1.702294 \\ \end{pmatrix}$&
$\begin{pmatrix} 0.04877305&0.02926383\\0.02926383& 0.04877305\\ \end{pmatrix}$\\
\hline
6&1/8&$\begin{pmatrix}-1.84197 \\ -1.84197 \\ \end{pmatrix}$&
$\begin{pmatrix}0.02167691&0.01300615\\0.01300615&0.02167691\\ \end{pmatrix}$\\
\hline
7&1/8&$\begin{pmatrix}-1.935086 \\ -1.935086 \\ \end{pmatrix}$&
$\begin{pmatrix}0.009634183&0.005780510\\0.005780510&0.009634183\\ \end{pmatrix}$\\
\hline
8&1/8&$\begin{pmatrix}-1.997164 \\ -1.997164 \\ \end{pmatrix}$&
$\begin{pmatrix}0.004281859&0.002569116\\0.002569116&0.004281859\\ \end{pmatrix}$\\
\hline
\end{tabular}
\label{tab:strong}
\end{table}
\pagebreak

\subsubsection{Kurtotic unimodal}

￼\begin{figure}[H]
\centering
\includegraphics[width=0.7\textwidth]{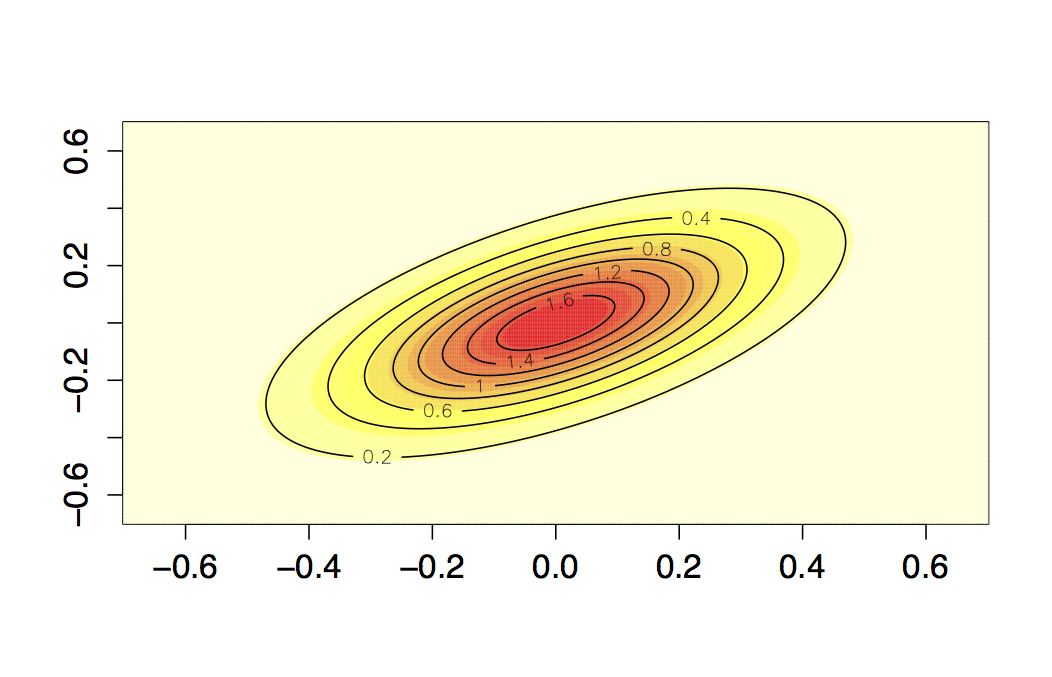}
\caption[Kurtotic unimodal distribution: Contour map]{Kurtotic unimodal distribution: Contour map denoting the density of a kurtotic unimodal distribution.}  
\end{figure}

\begin{table}[h]
\caption{Kurtotic unimodal distribution: Simulation settings.}
\center
\begin{tabular}{|l|c|c|c|}
\hline
Cluster&Prop&Mean&Covariance\\
&$\tau_g$&$(\mu_g)$&$(\Sigma_g)$\\ 
\hline
1&2/3&$\begin{pmatrix}0 \\ 0 \\ \end{pmatrix}$&
$\begin{pmatrix}1.25&0.75\\0.75&1.25\\ \end{pmatrix}$\\
\hline
2&1/3&$\begin{pmatrix}0 \\ 0 \\ \end{pmatrix}$&
$\begin{pmatrix}0.03952847&0.02371708\\0.02371708&0.03952847\\ \end{pmatrix}$\\
\hline
\end{tabular}
\label{tab:ku}
\end{table}
\pagebreak

\subsubsection{Outlier}

￼\begin{figure}[H]
\centering
\includegraphics[width=0.7\textwidth]{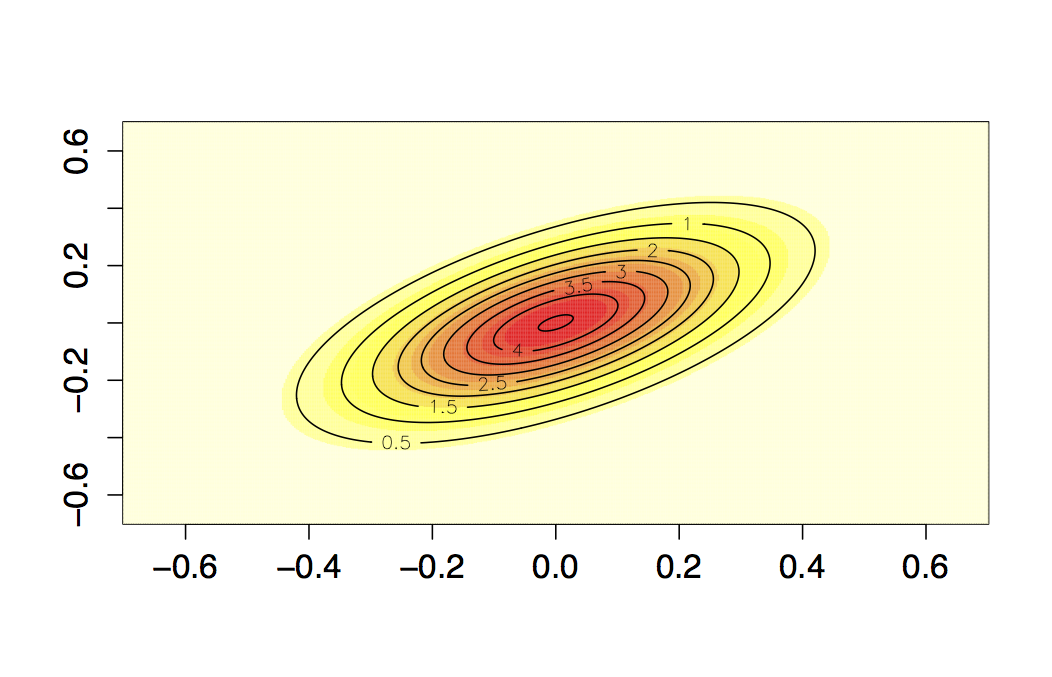}
\caption[Outlier distribution: Contour map]{Outlier distribution: Contour map denoting the density of a distribution with outliers.}   
\end{figure}

\begin{table}[h]
\caption{Outlier distribution: Simulation settings.}
\center
\begin{tabular}{|l|c|c|c|}
\hline
Cluster&Prop&Mean&Covariance\\
&$\tau_g$&$(\mu_g)$&$(\Sigma_g)$\\ 
\hline
1&1/10&$\begin{pmatrix}0 \\ 0 \\ \end{pmatrix}$&
$\begin{pmatrix}1.25&0.75\\0.75&1.25\\ \end{pmatrix}$\\
\hline
2&9/10&$\begin{pmatrix}0 \\ 0 \\ \end{pmatrix}$&
$\begin{pmatrix}0.03952847&0.02371708\\0.02371708&0.03952847\\ \end{pmatrix}$\\
\hline

\end{tabular}
\label{tab:outlier}
\end{table}
\pagebreak

\subsubsection{Bimodal}

￼\begin{figure}[H]
\centering
\includegraphics[width=0.7\textwidth]{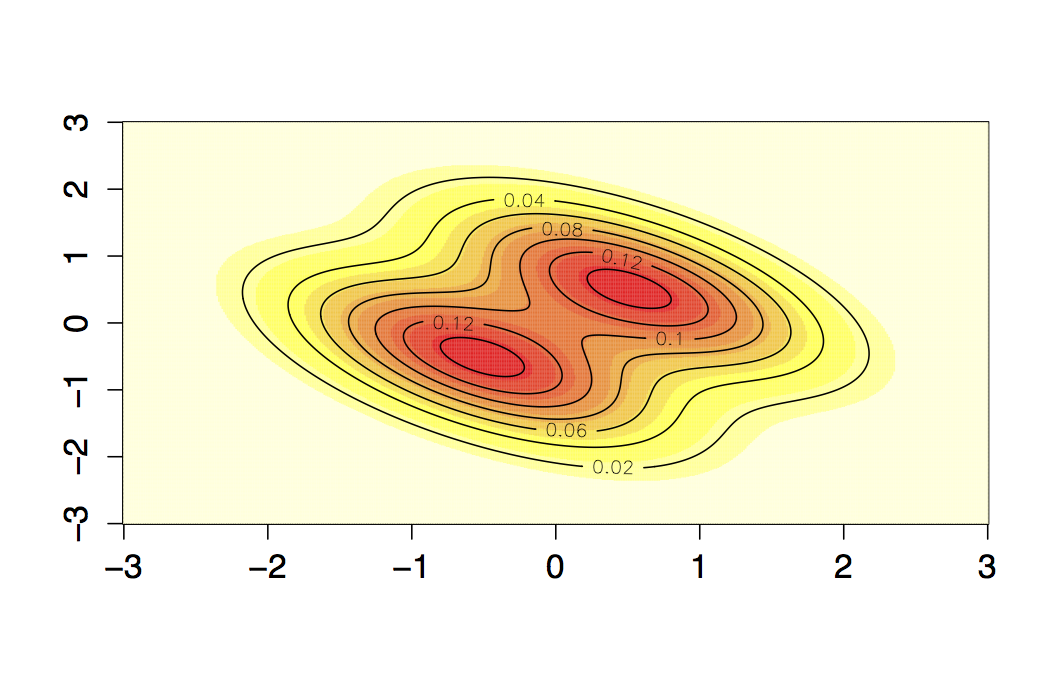}
\caption[Bimodal distribution: Contour map]{Bimodal distribution: Contour map denoting the density of a bimodal distribution.} 
\end{figure}

\begin{table}[h]
\caption{Bimodal Data: Simulation settings.}
\center
\begin{tabular}{|l|c|c|c|}
\hline
Cluster&Prop&Mean&Covariance\\
&$\tau_g$&$(\mu_g)$&$(\Sigma_g)$\\ 
\hline
1&1/2&$\begin{pmatrix}-0.5303301 \\ -0.5303301 \\ \end{pmatrix}$&
$\begin{pmatrix*}[r]0.6804138&-0.4082483\\-0.4082483&0.6804138\\ \end{pmatrix*}$\\
\hline
2&1/2&$\begin{pmatrix}0.5303301 \\ 0.5303301 \\ \end{pmatrix}$&
$\begin{pmatrix*}[r]0.6804138&-0.4082483\\-0.4082483&0.6804138\\ \end{pmatrix*}$\\
\hline

\end{tabular}
\label{tab:bimodal}
\end{table}
\pagebreak

\subsubsection{Separated bimodal}

￼\begin{figure}[H]
\centering
\includegraphics[width=0.7\textwidth]{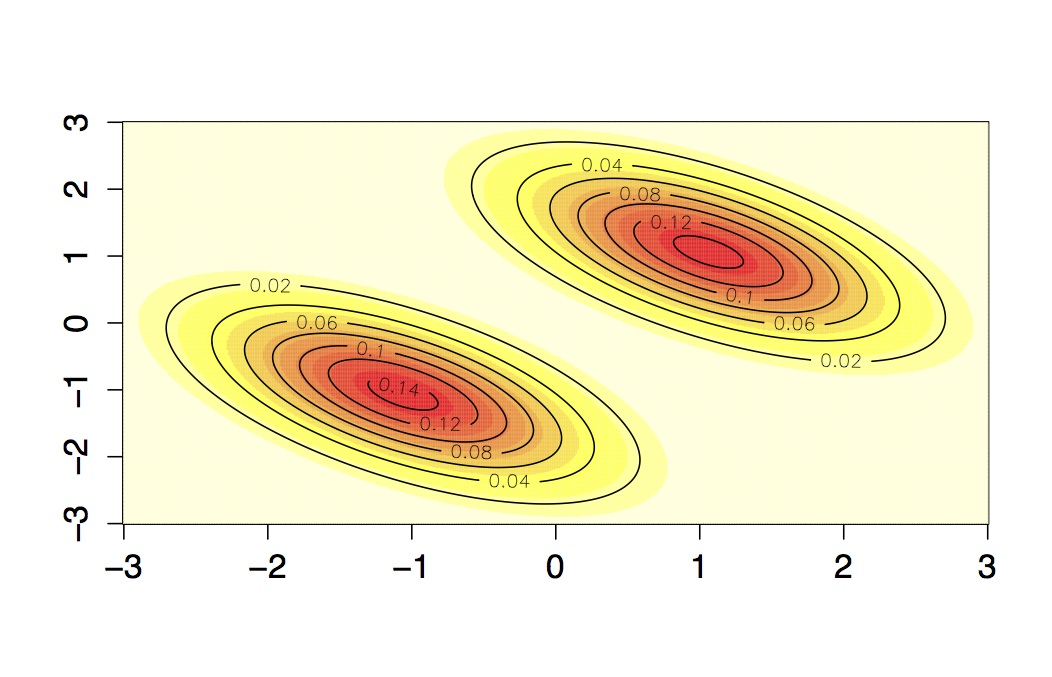}
\caption[Separated bimodal distribution: Contour map]{Separated bimodal distribution: Contour map denoting the density of a separated bimodal distribution.} 
\end{figure}

\begin{table}[h]
\caption{Separated bimodal distribution: Simulation settings.}
\center
\begin{tabular}{|l|c|c|c|}
\hline
Cluster&Prop&Mean&Covariance\\
&$\tau_g$&$(\mu_g)$&$(\Sigma_g)$\\ 
\hline
1&1/2&$\begin{pmatrix}-1.06066 \\ -1.06066 \\ \end{pmatrix}$&
$\begin{pmatrix*}[r]0.6804138&-0.4082483\\-0.4082483&0.6804138\\ \end{pmatrix*}$\\
\hline
2&1/2&$\begin{pmatrix}1.06066 \\1.06066 \\ \end{pmatrix}$&
$\begin{pmatrix*}[r]0.6804138&-0.4082483\\-0.4082483&0.6804138\\ \end{pmatrix*}$\\
\hline
\end{tabular}
\label{tab:sep}
\end{table}
\pagebreak

\subsubsection{Asymmetric bimodal}

￼\begin{figure}[H]
\centering
\includegraphics[width=0.7\textwidth]{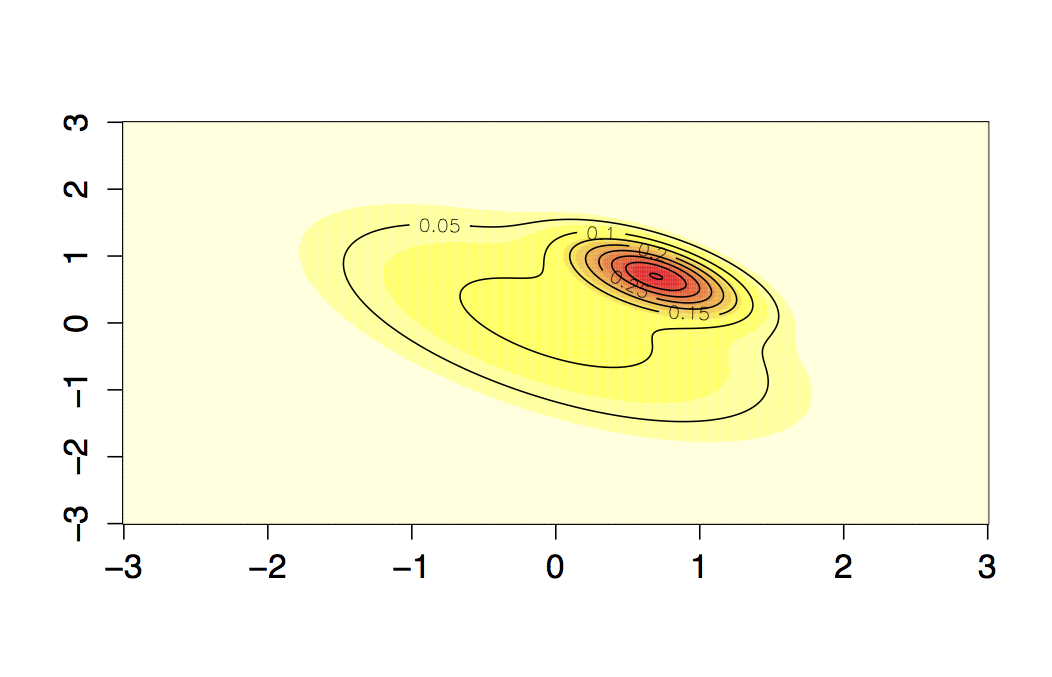}
\caption[Asymmetric bimodal distribution: Contour map]{Asymmetric bimodal distribution: Contour map denoting the density of an asymmetric bimodal distribution.} 
\end{figure}

\begin{table}[h]
\caption{Asymmetric bimodal distribution: Simulation settings.}
\center
\begin{tabular}{|l|c|c|c|}
\hline
Cluster&Prop&Mean&Covariance\\
&$\tau_g$&$(\mu_g)$&$(\Sigma_g)$\\ 
\hline
1&3/4&$\begin{pmatrix}0 \\ 0 \\ \end{pmatrix}$&
$\begin{pmatrix*}[r]1.25&-0.75\\-0.75&1.25\\ \end{pmatrix*}$\\
\hline
2&1/4&$\begin{pmatrix}0.7071068 \\ 0.7071068 \\ \end{pmatrix}$&
$\begin{pmatrix*}[r]0.13888889&-0.08333333\\-0.08333333&0.13888889\\ \end{pmatrix*}$\\
\hline
\end{tabular}
\label{tab:asymm}
\end{table}
\pagebreak

\subsubsection{Trimodal}

￼\begin{figure}[H]
\centering
\includegraphics[width=0.7\textwidth]{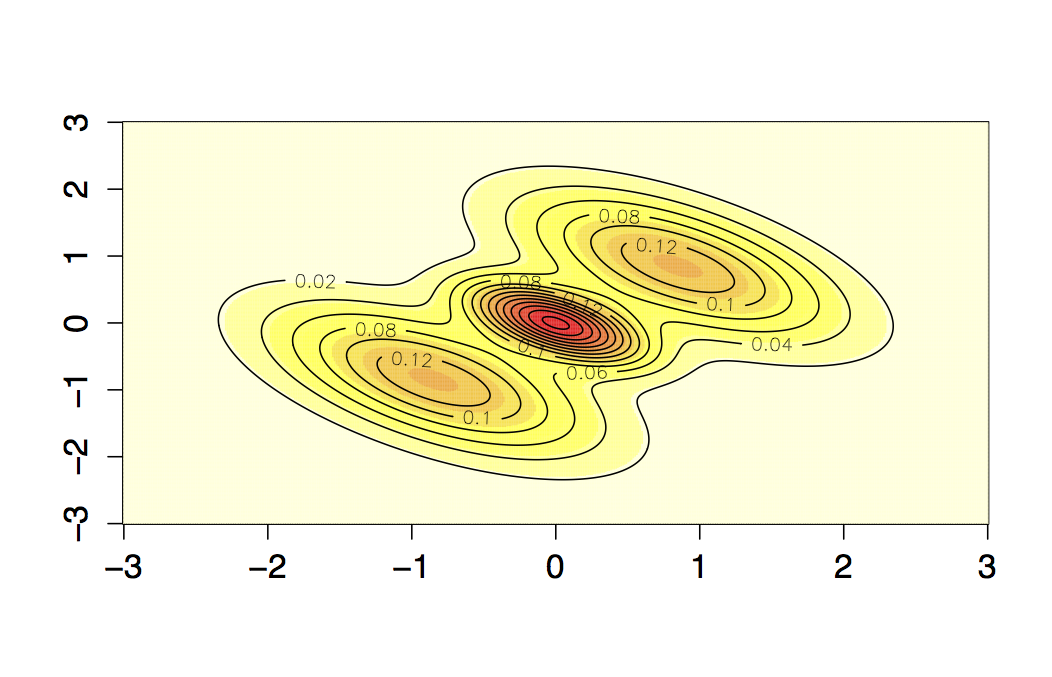}
\caption[Trimodal distribution: Contour map]{Trimodal distribution: Contour map denoting the density of a trimodal distribution.}   
\end{figure}

\begin{table}[h]
\caption{Trimodal distribution: Simulation settings.}
\center
\begin{tabular}{|l|c|c|c|}
\hline
Cluster&Prop&Mean&Covariance\\
&$\tau_g$&$(\mu_g)$&$(\Sigma_g)$\\ 
\hline
1&2/5&$\begin{pmatrix}-0.8485281 \\ -0.8485281 \\ \end{pmatrix}$&
$\begin{pmatrix*}[r]0.5809475&-0.3485685\\-0.3485685&0.5809475\\ \end{pmatrix*}$\\
\hline
2&2/5&$\begin{pmatrix}0.8485281 \\ 0.8485281 \\ \end{pmatrix}$&
$\begin{pmatrix*}[r]0.5809475&-0.3485685\\-0.3485685&0.5809475\\ \end{pmatrix*}$\\
\hline
3&1/5&$\begin{pmatrix}0 \\ 0 \\ \end{pmatrix}$&
$\begin{pmatrix*}[r]0.15625&-0.09375\\-0.09375&0.15625\\ \end{pmatrix*}$\\
\hline
\end{tabular}
\label{tab:trimodal}
\end{table}

\pagebreak
\subsubsection{Claw (Bart Simpson)}

￼\begin{figure}[H]
\centering
\includegraphics[width=0.7\textwidth]{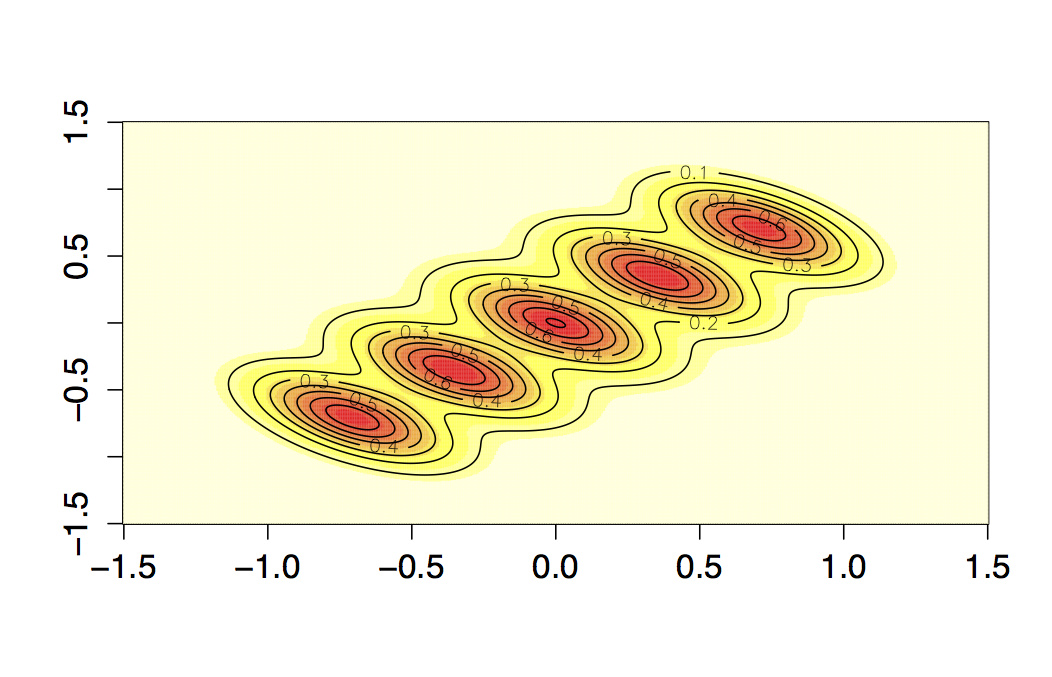}
\caption[Claw distribution: Contour map]{Claw distribution: Contour map denoting the density of the claw distribution.}   
\end{figure}

\begin{table}[h]
\caption{Claw distribution: Simulation settings.}
\center
\begin{tabular}{|l|c|c|c|}
\hline
Cluster&Prop&Mean&Covariance\\
&$\tau_g$&$(\mu_g)$&$(\Sigma_g)$\\ 
\hline
1&2/7&$\begin{pmatrix}0 \\ 0 \\ \end{pmatrix}$&
$\begin{pmatrix}0.625&0.375\\0.375&0.625\\ \end{pmatrix}$\\
\hline
2&1/7&$\begin{pmatrix}-0.7071068 \\ -0.7071068 \\ \end{pmatrix}$&
$\begin{pmatrix*}[r]0.03952847&-0.02371708\\-0.02371708&0.03952847\\ \end{pmatrix*}$\\
\hline
3&1/7&$\begin{pmatrix}-0.3535534 \\ -0.3535534 \\ \end{pmatrix}$&
$\begin{pmatrix*}[r]0.03952847&-0.02371708\\-0.02371708&0.03952847\\ \end{pmatrix*}$\\
\hline
4&1/7&$\begin{pmatrix}0 \\ 0 \\ \end{pmatrix}$&
$\begin{pmatrix*}[r]0.03952847&-0.02371708\\-0.02371708&0.03952847\\ \end{pmatrix*}$\\
\hline
5&1/7&$\begin{pmatrix}0.3535534 \\ 0.3535534 \\ \end{pmatrix}$&
$\begin{pmatrix*}[r]0.03952847&-0.02371708\\-0.02371708&0.03952847\\ \end{pmatrix*}$\\
\hline
6&1/7&$\begin{pmatrix}0.7071068 \\ 0.7071068 \\ \end{pmatrix}$&
$\begin{pmatrix*}[r]0.03952847&-0.02371708\\-0.02371708&0.03952847\\ \end{pmatrix*}$\\
\hline
\end{tabular}
\label{tab:claw}
\end{table}
\pagebreak
\subsection{Higher dimensional distribution settings}
\label{sec:highdimensional}
\subsubsection{Additional standard normal columns} 
For these results we merely added standard normal observations to the simulated bivariate clusters. We did not differentiate between the clusters as cluster membership is not important in this setting.
\subsubsection{Three- and six-dimensional extensions}
 \begin{table}[h]
\caption[3D Bimodal Data: Simulation settings]{ 3D Bimodal Data: Settings used to simulate the density of a 3D bimodal distribution.Note that displacement means the distance from the origin - in this case along the line $y=x$. Separation as described in the results tables is the separation of the group centres.}
\center
\begin{tabular}{|l|l|c|c|c|}
\hline
Cluster&Separation&Prop&Mean&Covariance\\
&&$\tau_g$&$(\mu_g)$&$(\Sigma_g)$\\ 
\hline
1&1.5&1/2&$\begin{pmatrix}-0.5303301 \\ -0.5303301 \\0.0\\ \end{pmatrix}$&
$\begin{pmatrix*}[r]1.5&-0.5&0.0\\-0.5&1.5&0.0\\0.0&0.0&0.5\\ \end{pmatrix*}$\\
\hline
2&&1/2&$\begin{pmatrix}0.5303301 \\ 0.5303301 \\ 0.0\\ \end{pmatrix}$&
$\begin{pmatrix*}[r]1.5&-0.5&0.0\\-0.5&1.5&0.0\\0.0&0.0&0.5\\ \end{pmatrix*}$\\
\hline
\hline
1&3&1/2&$\begin{pmatrix}-1.06066 \\ -1.06066 \\0.0\\ \end{pmatrix}$&
$\begin{pmatrix*}[r]1.5&-0.5&0.0\\-0.5&1.5&0.0\\0.0&0.0&0.5\\ \end{pmatrix*}$\\
\hline
2&&1/2&$\begin{pmatrix}1.06066 \\ 1.06066 \\ 0.0\\ \end{pmatrix}$&
$\begin{pmatrix*}[r]1.5&-0.5&0.0\\-0.5&1.5&0.0\\0.0&0.0&0.5\\ \end{pmatrix*}$\\
\hline
\hline
1&5&1/2&$\begin{pmatrix}-1.767767 \\ -1.767767 \\0.0\\ \end{pmatrix}$&
$\begin{pmatrix*}[r]1.5&-0.5&0.0\\-0.5&1.5&0.0\\0.0&0.0&0.5\\ \end{pmatrix*}$\\
\hline
2&&1/2&$\begin{pmatrix}1.767767 \\ 1.767767 \\ 0.0\\ \end{pmatrix}$&
$\begin{pmatrix*}[r]1.5&-0.5&0.0\\-0.5&1.5&0.0\\0.0&0.0&0.5\\ \end{pmatrix*}$\\
\hline

\end{tabular}
\label{tab:bimodal3}
\end{table}

 \begin{table}[h]
\caption[6D Bimodal Data: Simulation settings]{ 6D Bimodal Data: Settings used to simulate the density of a 6D bimodal distribution. Note that displacement means the distance from the origin - along the line $x_1=x_2$, as before. Separation as described in the results tables is the separation of the group centres.}
\center
\begin{tabular}{|l|l|c|c|c|}
\hline
Cluster&Displacement&Prop&Mean&Covariance\\
&&$\tau_g$&$(\mu_g)$&$(\Sigma_g)$\\ 
\hline
1&0.75&1/2&$\begin{pmatrix}-0.5303301 \\ -0.5303301 \\0.0\\0.0\\0.0\\0.0\\ \end{pmatrix}$&
$\begin{pmatrix*}[r]3.0&1.0&0&0&0&0\\1.0&3.0&0&0&0&0\\0&0&1&0&0&0\\ 0&0&0&1&0&0\\ 0&0&0&0 &0.5 &0\\0&0&0&0&0 &0.25\\\end{pmatrix*}$\\
\hline
2&&1/2&$\begin{pmatrix}0.5303301 \\ 0.5303301 \\0.0\\0.0\\0.0\\0.0\\ \end{pmatrix}$&
$\begin{pmatrix*}[r]3.0&1.0&0&0&0&0\\1.0&3.0&0&0&0&0\\0&0&1&0&0&0\\ 0&0&0&1&0&0\\ 0&0&0&0 &0.5 &0\\0&0&0&0&0 &0.25\\\end{pmatrix*}$\\
\hline
1&1.5&1/2&$\begin{pmatrix}-1.06066 \\ -1.06066 \\0.0\\0.0\\0.0\\0.0\\ \end{pmatrix}$&
$\begin{pmatrix*}[r]3.0&1.0&0&0&0&0\\1.0&3.0&0&0&0&0\\0&0&1&0&0&0\\ 0&0&0&1&0&0\\ 0&0&0&0 &0.5 &0\\0&0&0&0&0 &0.25\\\end{pmatrix*}$\\
\hline
2&&1/2&$\begin{pmatrix}1.06066 \\ 1.06066 \\0.0\\0.0\\0.0\\0.0\\ \end{pmatrix}$&
$\begin{pmatrix*}[r]3.0&1.0&0&0&0&0\\1.0&3.0&0&0&0&0\\0&0&1&0&0&0\\ 0&0&0&1&0&0\\ 0&0&0&0 &0.5 &0\\0&0&0&0&0 &0.25\\\end{pmatrix*}$\\
\hline
1&2.5&1/2&$\begin{pmatrix}-1.767767 \\ -1.767767 \\0.0\\0.0\\0.0\\0.0\\ \end{pmatrix}$&
$\begin{pmatrix*}[r]3.0&1.0&0&0&0&0\\1.0&3.0&0&0&0&0\\0&0&1&0&0&0\\ 0&0&0&1&0&0\\ 0&0&0&0 &0.5 &0\\0&0&0&0&0 &0.25\\\end{pmatrix*}$\\
\hline
2&&1/2&$\begin{pmatrix}1.767767 \\ 1.767767 \\0.0\\0.0\\0.0\\0.0\\ \end{pmatrix}$&
$\begin{pmatrix*}[r]3.0&1.0&0&0&0&0\\1.0&3.0&0&0&0&0\\0&0&1&0&0&0\\ 0&0&0&1&0&0\\ 0&0&0&0 &0.5 &0\\0&0&0&0&0 &0.25\\\end{pmatrix*}$\\
\hline
\end{tabular}
\label{tab:bimodal6}
\end{table}
\end{document}